# Classical Thought in Newton's General Scholium

*Karin Verelst**

*karin.verelst@vub.be*

*From Thales to Paul by way of the Stoics*

B.J.T. Dobbs

## Abstract

*Isaac Newton, in popular imagination the Ur-scientist, was an outstanding humanist scholar. His researches on, among others, ancient philosophy, are thorough and appear to be connected to and fit within his larger philosophical and theological agenda. It is therefore relevant to take a closer look at Newton's intellectual choices, at how and why precisely he would occupy himself with specific text-sources, and how this interest fits into the larger picture of his scientific and intellectual endeavours. In what follows, we shall follow Newton into his study and look over his shoulder while reading compendia and original source-texts in his personal library at Cambridge, meticulously investigating and comparing fragments and commentaries, and carefully keeping track in private notes of how they support his own developing ideas. Indeed, Newton was convinced that precursors to his own insights and discoveries were present already in Antiquity, even before the Greeks, in ancient Egypt, and he puts a lot of time and effort into making the point, especially, and not incidentially, in the period between the first and the second edition of the* Principia. *A clear understanding of his reading of the classic sources therefore matters to our understanding of its content and gestation process. In what follows we will confine ourselves to the classical legacy, and investigate Newton's intellectual intercourse with it.*

## Introduction[1]

Newton was much more than what we would call today an 'exact scientist'. His wide-ranging intellectual backgound and his keen and active interest in fields like theology and philology shape his research projects as much as do mathematics and mechanics, all be it in a less obvious way. Newton was competent in these areas as much as in physics, thanks to his university education and his own sustained efforts to stay abreast of the latest contemporary developments.[2] Intellectual training at university in Newton's time consisted mainly in the study of the humanities, i.e., of philosophical and literary traditions reaching back millenia, and which continued to be heavily influenced by ancient Greek philosophy, especially Artistotle.[3] From his earliest days as a student, Newton engages in an avid discussion with these giants of the past which he respects, but from

* Center Leo Apostel (CLEA), the interdisciplinary research institute of the Vrije Universiteit Brussel. A part of this research was carried out while I was NIAS Fellow at the Dutch Royal Academy, 2016-2017.

[1] I owe thanks to the hospitality of Scott Mandelbrote at Cambridge (spring 2014), to the generosity of the curators at Wren Library in Cambridge in that same year, to the staff of the Library of the Royal Society in London in 2018, and Stephen Snobelen and Steffen Ducheyne, for Ms. material collected and generously shared with me. All those mentioned contributed to the development of the ideas presented here through inspiring discussions and critical comments. Remaining deficiencies are only my responsibiity.

[2] This was not uncommon for an intellectual of standing in these days, even when his business was mainly what we today call 'science'. An interesting other example is discussed in D. Levitin, "The Experimentalist as Humanist: Robert Boyle on the History of Philosophy", *Annals of Science*, 71, 2 (2014): 149-182.

[3] Although slowly evolving in a more humanist direction, university education in Newton's time was still steeply rooted in the Scholastic tradition, especially at the undergraduate level. In fact, in the seventeenth century, Aristotelianism underwent an intellectual rejuvenation in the universities. See M. Feingold, "Aristotle and the Englsih universities in the seventeenth century: a reevaluation", in: *European Universities in the Age of Reformation and Counter-Reformation*, H. Robinson-Hammerstein (ed.), Dublin (1998): 135-148. The curriculum to which Newton was exposed in his early student years (of which we are well informed through the commonplace notebook he held from 1661 till 1665), was based on a 'modern' version of the Peripatetic legacy. For a description of Newton's notes from Greek and Latin sources, which fill the major part of the notebook, see J.E. McGuire and M. Tamny, *Certain Philosophical Questions: Newton's Trinity Notebook*, Cambride (1983): 15-20; Also S. Ducheyne, "Newton's Training in the Aristotelian Textbook Tradition: From Effects to Causes and Back", *History of science*, 43 (2005). Some authors want to downplay the impact of Scholastic pedagogy on young Newton, claiming it was "superficial"; J.Z. Buchwald and M. Feingold, *Newton and the Origin of Civilisation*, Princeton (2012): 14–15. I disagree. Even when reluctantly, Newton was trained thoroughly in the understanding and use of the core metaphysical concepts of the tradition, as well as in the use of the analytical method of *disputatio*, as is clear from many of his manuscripts. We will have the occasion to see some examples of both. "Peripateticism, in whatever propaedeutic form, was the earliest contact they had as individuals with serious philosophical and scientific concerns. However unsatisfying it became for some of them, at least it comprised a rigorously organised body of doctrine that included a systematic interpretation of the diversities of nature. It showed the thoughtful student of nature that an intelligible and comprehensive account of natural phenomena was a *prima facie* possibility." R. Ariew and A. Gabbey, "The Scholastic Background", in: The Cambridge History of Seventeenth-Century Philosophy, D. Garber and M. Ayers (eds.), Cambridge (1998): 425. For the more general context: James Hannam, *Teaching natural philosophy and mathematics at Oxford and Cambridge, 1500–1700*, PhD thesis, University of Cambridge (2008). For a historical critical study of influential precursors for today's science in the late scholastic period, see A. Maier, *Studien zur Naturphilosophie der Spätscholastik*, 5 vols., Rome (1949–1958), especially vol. 2: *An der Grenze von Scholastik und Naturwissenschaft* (1952).

which he nevertheless wishes to break free.[4] But his own research throws him back again and again onto the questions they raised. Decades of research of the papers in his legacy thus reveal a very different and quite surprising side of his intellectual persona: the humanist scholar who delves deep into sources partaining to history, philology, theology, and even alchemy[5]. Newton was not just studying these subjects out of mere leisurely interest, but approached them with a methodical thoroughness and a zeal comparable to his research in natural philosophy.[6] His interest, moreover, extends over decades, spanning the whole of his intellectual career. We find him writing lengthy treatises on the results of his investigations into ancient sources which almost never got published, but which were shared by him to a few trusted interlocutors and friends and then distributed among selected members of his inner circle. Thus they could exert a considerable, al be it indirect, influence on the intellectual and theological debates of his time.

An intriguiging aspect of this interest in humanist scholarship is Newton's aspiration to connect his own theories directly to the oldest intellectual sources of our civilisation, while framing himself as the rediscoverer of lost wisdom from the distant past.[7] Again this is a characteristic feature of his intellectual persona from the beginning till the very end. It occasionally surfaces in the *Principia* itself, in the mathematical sections as well as in the text which is at the focus of attention in this volume, the General Scholium.[8]

[4] As is clear from the motto he choses to head his first personal research efforts, "Amicus Plato amicus Aristoteles magis amica veritas", McGuire and Tamny, *Trinity Notebook*, 336.
[5] The researxch on Newton's alchemy has been pineered by B.J.T. Dobbs, *The foundations of Newton's Alchemy, or The hunting of the Greene Lyon*, (Cambridge, 1975). Also K. Figala, "Newton's alchemy", in I.B. Cohen and G.E. Smith (eds.), *The Cambridge Companion to Newton* (Cambridge, 2002) 370-386.
[6] Recent books on the different aspects of Newton's intellectual persona: R. Iliffe, *Priest of Nature*, Oxford, 2017; Janiak *Newton as Philosopher* (Cambridge, 2008); Buchwald and Feingold, *Origins of Civilisation*; and of course Dobbs's classic *The Janus Face of Genius. The role of Alchemy in Newton's Thought*, (Cambridge, 1992).
[7] On the scope and import of Newton's self-styled indebtedness to a presumed *prisca sapientia*, scholarly opinion tends to diverge. The idea was pioneered by J. E. McGuire and P. M. Rattansi, "Newton and the 1 of Pan", *Notes and records of the Royal Society of London,* xxi (1966): 108-43. Its scope and depth was considerably enlarged by B.J.T. Dobbs, "Newton and Stoicism", *The Southern Journal of Philosophy*, vol. 23 (suppl.) (1985): 112: "He also had a deep-seated belief in an ancient wisdom (*prisca sapientia*) that stretched back through the ages, through the earliest sages of mankind ultimately to Adam to whom God had given perfect knowledge." See also Dobbs, *Green Lyon*; —, *Janus Faces.* A critical reply by Casini in his edition of the Classical Scholia (see below for more details). Recently also more criticism in D. Levitin, *Ancient Wisdom in the Age of the New Science* (Cambridge, 2015). A balanced assessment in Iliffe, *Priest*, especially Chapter 6.
[8] E.g., Book I, Sec. V, Lemma XIX, cor. 2 [148-149 in the *variorum*-edition], where Newton explicitly states that he solved Pappus's Problem by means of methods akin to those of the ancient geometers: "*Non calculus, sed compositio geometrica:* this maxim agrees with the *Principia* as a whole, at least if it is allowed that the geometry of the ancients can include quantities generated by motion and metamorphoses ending in an 'ultimate state'", so F. De Gandt, *Force and Geometry in Newton's Principia*, transl. C; Wilson (Princeton, 1995): 242-244 (who refers, however, to "Lemma 18"). Considerably more elaborated, N. Guicciardini, *Reading the Principia. The debate on Newton's mathematical methods for Natural Philosophy from 1687 to 1736* (Cambridge, 1999), sec. 2.3.1 'The Ancients versus the Moderns', 27-32; Ch. 4: 'Newton: between tradition and innovation', 99-108, and especially the section "Prisca geometria", 101-104: "Newton considered himself a rediscoverer of an ancient knowledge and came to attribute to the Ancients the doctrines of atoms, of the void, of the

However, classical sources occupied Newton far more than might appear from such scarcely published mentions or references. He possessed a rich personal library, containing many of the top level editions of the classics of his day, often showing the marks of intensive use.[9] From his student days onwards, Newton lavishly quotes and discusses in his manuscripts tropes and topics of classical origin, investigating them methodologically in order to prove that "the Ancients" (*Veteres*) in many essential respects were indeed precursors to his own cosmological and theological ideas.[10] An eloquent and well-known example of this fact are the so-called Classical Scholia from the manuscript legacy of David Gregory, a series of commentaries to propositions IV - IX of book III of the *Principia*, and destined to be integrated in the second edition Newton was envisaging and preparing during the 1690s.[11] We know moreover that these notes relate directly to the General Scholium, not only because of their contents, but also because they are joined by their compiler to some manuscript drafts which found their way into the General Scholium in either its 1713, or its 1726 printed version.[12] In what follows, I investigate in detail what these two texts — General Scholium and Classical Scholia — tell us about Newton's methodological approach towards the classical sources and about planetary nature of the Earth and of universal gravitation. Newton not only believed that his *philosophia naturalis* was a rediscovery of ancient philosophy, he also stated that his *principia mathematica* were a rediscovery of ancient geometrical methods", *o.c.*, 102.

[9] J. Harrison, *The Library of Isaac Newton,* Cambridge (1978). An updated version on the Newton Project website: http://www.newtonproject.ox.ac.uk/his-library/books-in-newtons-library.

Examples of such usage (like dog-ears) in S. Mandelbrote, *Footprints Of The Lion: Isaac Newton At Work: Exhibition At Cambridge University Library*, 9 October 2001-23 March (2002).

[10] Newton considered for a while publishing *A Treatise of the System of the Wor*ld, a popularised version of the future Book III of the *Principia*, around 1685. The introduction discusses ancient philosophies and religious practices that refer, according to him, to heliocentrism. Dobbs, *Janus Faces*, 186-187.

[11] An autograph in Gregory's Memorandum of 1694; Greg MS 247 ff 6-14v, Library of the Royal Society. For Gregory's Memoranda, see *The Correspondence of Isaac Newton*, H. W. Turnbull et al. (eds.), 7 vols, Cambridge (1959-1978), especially the Memorandum of 5-7 May, 1694, 3: 446. Discussion: A. Rupert Hall, "Correcting the Principia", *Osiris*, 13 (1958): 291-326; Cohen, "'Quantum in se est'. Newton's Concept of Inertia in Relation to Descartes and Lucretius", *Notes and Records of the Royal Society of London*, 19, 2 (1964): 131-155. First edition: P. Casini, "The Classical Scholia", *History of science* 22, (1984) (translation of the Italian original, 1981). Some corrections to this edition in R. De Smet and K. Verelst, "Platonic and Stoic Legacy", *History of Science*, 39 (2001): 1-30. For a revised and extensively commented edition: V. Schüller, "Newton's Scholia from David Gregory's Estate on the Propositions IV to IX book III of his Principia", in: *Between Leibniz, Newton and Kant*, W. Lefèvre (ed.), Springer, Dordrecht (2001): 213-265. This is an English translation of a working paper for the Max Planck Instiitute: V. Schüller, *Newton's Scholia aus David Gregorys Nachlaß zu den Propositionen IV- IX Buch III seiner* Principia, Max-Planck-Institut für Wissenschaftsgeschichte, Preprint 144 (2000). Schüller also transcribed Gregory's introduction to his *Astronomiae physicae & geometricae elementa from* 1702. The appendix contains facsimile reproductions of the manuscript (Royal Society Ms. 247: ff. 6-14).

[12] "It ends with three paragraphs, the first and third of which appear to be an early draft of the general Scholium at the end of the second edition of the *Principia* (1713), while the second paragraph formed the substance of a passage which is only found in the third edition (1726)." So Turnbull, *Correspondence* 3, 339. For a detailed study of the different editorial phases the General Scholium ran before its publication, see S. Ducheyne, "Some Notes on Newton's Published and Unpublished Endeavours", *Lias: Sources and Documents Relating to the Early Modern History of Ideas*, vol. 33, no. 3n (2006): 223-274.

the reasons which made him incessantly search for ancient support of his own basic mathematical, philosophical and theological positions. The General Scholium moreover occupies a unique place because it allows us to peak behind the corner and witness how the two strands — natural philosophy and theology — come together on the most fundamental level of Newton's thought.

Newton's preoccupation with ancient sources and literature has been the subject of in-depth research, pioneered by Westfall, Rattansi, Dobbs, Cohen and Casini. Its relevance has been further confirmed by different scholars establishing the conceptually close ties between Newton's published work and the pile of manuscripts and notes on philosophical, theological and alchemical subject-matters that fill the vaults of Cambridge University Library.[13] But the scholarship continues to be at variance with the way in which this fact has to be understood: are the common topics treated in these different fields of interest related or not? The positions depend on whether a 'holistic' or a 'disciplinary-segregational' point of view with respect to the intellectual origins of Newton's work is taken.[14] Rob Iliffe summarises the debate:

> The idea that apparently disparate parts of his writings are somehow connected was to some extent a response to the positivist emphasis of earlier Newton scholarship, but more generally, it was based on the metaphysical presumption that the individual 'Isaac Newton' was the undifferentiated author of a group of writings that were all coherent or unified at some level. (...) I suggest that from the very beginning of his researches, Newton shaped his own work according to distinctions between what was appropriate to the distinct disciplinary traditions of natural philosophy and mixed mathematics. (...) Although his writing in these fields ostensibly concerned identical phenomena (such as gravitation), for the most part they were fundamentally incompatible and there was little if any interaction or connection between them.[15]

[13] And other archives, like the Yahuda Archive in Jerusalem. Their collections are published electronically by https://www.newtonproject.ox.ac.uk/

[14] The terminology stems from Rob Iliffe (ft 14). The 'holistic' point of view is defended by Dobbs, *Janus faces*, and J.E. Force, "Newton's God of Dominion Newton's God of Dominion: The Unity of Newton's Theological, Scientific, and Political Thought, in: *Essays on the Context, Nature, and Influence of Isaac Newton's Theology*, James E. Force and Richard H. Popkin (eds.), *Archives Internationales D'Histoire Des Idées*, vol. 129 (1990): 75-102; James E. Force, "The nature of Newton's 'holy alliance' between science and religion: from the scientific revolution to Newton (and back again)", in *Rethinking the Scientific Revolution*, Margaret J. Osler (ed.), Cambridge (2000): 247-70, and more recently by Stephen D. Snobelen, "God of Gods, and Lord of Lords. The Theology of Isaac Newton's General Scholium to the Principia", *Osiris*, 16 (2001). The 'segregational' point of view has been defended by R.S. Westfall, "The Scientific Revolution Reasserted", in Osler, *Scientific Revolution*, 41-59. It is also prominent in Buchwald and Feingold, *Civilisation*, 139-147, and in Levitin, *Ancient Wisdom*, 433-446.

[15] R. Iliffe, "Abstract considerations: disciplines and the incoherence of Newton's natural philosophy", *Stud.*

This, of course, only makes sense if it be valid not only for Newton's dealings with philosophy and mathematics, but also theology, alchemy and all the rest. B.J.T. Dobbs, as is well known, defends the deep interconnectedness of Newton's different intellectual endeavours.[16] From the point of view taken in this contribtion it does not really matter what Newton's deepest motivations were. Given the overwhelming amount of direct textual evidence to the contrary as well as first hand witnesses, it would seem strange to simply discard the idea of any intrinsic link between, e.g., Newton's theology and his natural/experimental philosophy. Newton's own position with respect to this evolved, as Cohen makes clear:

> Thus Newton's statements about not 'feigning' hypotheses and about discussions of God in experimental philosophy were both additions to the text of the Scholium Generale, and both were tacked on at the same time! It would seem that Newton, once having decided (...) to append a strong statement that hypotheses of whatever kind have no place in experimental philosophy, felt he must add a sentence to show that God was not to be considered an hypothesis in experimental philosophy — because one could discuss him 'ex phaenomenis'.[17]

The key issue is the nature of the gravitational force which governs the motion of bodies in the universe, which translates theologically into the relation between God and the world.[18] I agree with Iliffe in his recent book on Newton's religious worlds that the major underlying themes of Newton's thought are already present right from the start, as can be gleaned from his earliest notes in the so-called Trinity Notebook.[19] The questions raised there resurface powerfully in the *De Gravitatione* manuscript two decades later.[20] And it

*Hist. Phil. Sci.* 35 (2004): 428-430.

[16] Dobbs, *Janus Face*; — *Stoicism*.

[17] Cohen, *Introduction to Newton's Principia*, 244. On possible reasons for this 'radicalisation', see A.E. Shapiro, "Newton's Experimental Philosophy", *Early Science and Medicine* 9, 3 (2004): 168-217. Ori Belkind's observation may be appropriate here: "we should distinguish between Newton the scientist and Newton the rhetorician. Newton the scientist made conjectures and hypothesized that the motions of the planets are governed by a force of gravitation obeying the inverse-square law. Newton the rhetorician claimed to have followed a strict inductive method." O. Belkind, Newton's scientific method and the universal law of gravitation, in: A. Janiak and E. Schliesser, *Interpreting Newton. Critical Essays* (Cambridge, 2012), 142.

[18] "Newton's own shifting conceptions of the ether reflect the very issues discussed here, so his pre- and post-*Principia* discussions of the properties of the ether differ fundamentally from one another." Janiak, *Philoosopher*, 100, ft. 21.

[19] McGuire and Tamny, *Trinity Notebook*.

[20] Discussion of this important manuscript falls out of the scope of this paper. A few references by way of oriëntation: published in A. Rupert Hall and Marie Boas Hall, *Unpublished Scientific Papers of Isaac Newton*, Cambridge (1978 [1962]) 89-156. J.A. Ruffner, "Newton's De gravitatione: a review and reassessment", *Archive for the History of the Exact Sciences*, 66, 2 (2012): 241-264. For some of its theological implications: H.

will take another two decades until the second edition of the *Principia* and the General Scholium for them to fully ripen and come out into the open. The connection between the development of ideas in Newton's natural philosophy and his theology is rather self-evident throughout, and dates back well before the period of conception of the *Principia*.[21] Though Newton's primary religious concerns need not have been explicitly "metaphysical",[22] given his presumed "explanatory agnosticism"[23], his statements are often deliberatedly ambigous[24] — this is certainly true of the General Scholium, the text that is our main concern here.[25] Moreover, the position "not to be metaphysical" is in itself eminently metaphysical, even in the disguise of "method".[26] Al least since Kuhn it is known that constructing an experiment implies constructing the at least partial worldview in which the questions raised by it can be asked and the potential answers to them make sense, i.e., it requires patterns of cognitive and conceptual expectation that preceed the outcome of the experiment.[27] That is why Koyré is right in his claim that the Scientific Revolution of the seventeenth century was above all a matter of metaphysics rather than

---

Kochiras, "By y$^{e}$ divine arm: God and substance in *De gravitatione*", Religious Studies 49, 3 (2013): 327-256. A recent paper by Levitin offers some new perspectives: D. Levitin, "Isaac Newton's 'De gravitatione et aequipondio fluidorum': its purpose in historical context", *Annals of Science* (2021).

21 This is clear from the fact that Newton references them extensively in the introduction to his *Treatise of the System of the World*, published postumously in 1728, but written well before the publication of the Principia, as we know from Halley's letters. S.P. Rigaud, *Historical Essay on the First Publication of Sir Isaac Newton's Principia*, Oxford University Press, Oxford (1838): 77-80.

22 F. Manuel, *The religion of Isaac Newton*, Clarendon, Oxford (1974).

23 Levitin, *Ancient Wisdom*, 436.

24 Blatantly so in his third letter to Bentley: "Gravity must be caused by an Agent acting constantly according to certain laws; but whether this Agent be material or immaterial, I have left to the connsideration of my readers." I. Bernard Cohen and Robert E. Schofield, *Isaac Newton's Papers and Letters on Natural Philosophy*, Cambridge (1958): 303.

25 As e.g. Janiak rightfully notes, "Newton then adds: 'It is agreed that the supreme God necessarily exists, and by the same necessity he is always and everywhere"'. Therefore, action on a distance cannot exist: "From Newton's point of view, then, God in fact never acts at a distance on any object, at any time in the history of the world." This bears directly on his understanding of how gravity works: "(…) it seems clear that when he contemplates the idea that God might be the relevant mediating element he is not contemplating the idea that bodies act at a distance on one another. Instead, God acts locally and directly on any object at any time." A. Janiak, *Newton as Philosopher*, Cambridge (2008): 37-39. Also J. E. McGuire, 'The fate of the date: the theology of Newton's *Principia* revisited', in *Rethinking the Scientific Revolution* Margaret J. Osler (ed.) (Cambridge, 2000): 271–95, 279–93.

26 The classic argument is made in E.A. Burtt, *The metaphysical foundations of modern physical science* (New York,, 1999 [1924]): 227-229. See also J. Ladyman, "Science, Metaphysics and Method", *Philosophical Studies*, 160 (2012): 31-51, 2012. For Newton from this perspective: S. Ducheyne, *The Main Business of Natural Philosophy: Isaac Newton's Natural-Philosophical Methodology*, Springer (New York, 2012).

27 I. Hacking, *Representing and Intervening* (Cambridge, 1983), especially Part B. This was the zest in the controversy with Hooke: "Hooke's methodological message was that the experiments warranted neither Newton's "hypothesis" nor even his own, for he could explain the phenomena according to two or three *other* theories equally well." M. Feingold, "Mathematicians and Naturalists: Sir Isaac Newton and the Royal Society", in: *Isaac Newton's Natural Philosophy*, J.Z. Buchwald and I.B. Cohen (eds.), (Cambridge, Mass., 2001): 84.

of empirical advance.[28] It might be worthwhile to remind the reader that whatever we believe the "Scientific Revolution" to have been, it began with a *discours de la methode*.

This sheds light on the issue of Newton's theological positions, specificially his antitrinitarianism. When God is 'substantialised' in his relation to his Son as is the case in the Nicene Creed[29], it is logically impossible for him to entertain the relation to the world Newton's natural philosophy requires.[30] From the Trinitarian point of view, the connection between the Christology and the ontology is immediate and inevitable.[31] This has of course consequences for those who reject the position.

> (...) the claims in the General Scholium are about God's operation in the world, not about his relationship to his son (...) That Newton's antitrinitarianism was connected to his natural philosophical statements does not mean that each of those natural philosophical statements contained an esoteric antitrinitarian message.[32]

Indeed, but it does not imply the opposite either. The fact of the matter is that the history of the reception of Christianity in late antiquity cannot be separated from its theological translation into Greek philosophy,[33] which in turn cannot be separated from the great theological debates of the fourth to sixth centuries that later gave rise to the schism of 1054. So yes, the operation of God in the world and the relation He entertains to his son are closely interconnected, and Newton was well aware of that.[34] Hence, even indepently from his own theological stance, his caution on the subject is completely understandable and commandable, and had nothing to do with a lack of interest in nor the relevance of the philosophical aspects of his theology. This is abundantly clear from manuscripts like the *Paradoxical questions regarding Athanasius*,[35] written in the early 1690s,[36] which starts out with a critical assessment of the historiography of the controversy between Arian and Athanasius, but ends with a disputation concerning *homousia*, the term the Nicene Creed uses to describe the relation between the Father and the Son.[37] This document dates from exactly the same period in which Newton worked intensely on his Classical Scholia.

---

[28] As paraphrased by P. Dear, *Discipline and Experience. The Mathematical Way in the Scientific Revolution*, Chicago (1995): 12. See also Peter R. Anstey, John A. Schuster (eds.), *The Science of Nature in the Seventeenth Century. Patterns of Change in Early Modern Philosophy*, Springer (Dordrecht, 2005).

[29] J.H. Leith, *Creeds of the Churches: A Reader in Christian Doctrine* (Westminster, 1982): 28–31.

[30] In the period after the publication of the second edition of the *Principia* (with the General Scholium), which is as well the period of the last great controversy of his life (the Leibniz-Clarke correspondence), Newton drafts a text (Keynes MS 8) which is his personal alternative for the (Trinitarian) Nicene Creed. The fact that he uses explicitly a pre-Trinitarian text as starting point (the Apostle's Creed), gives us, well, a clue as to the direction in which to look for an interpretation. The first article sounds like a theological shorthand for the content of the general scholium: "There is one God the Father, everliving, omnipresent, omniscient, almighty, the Maker of Heaven and Earth, and one Mediator between God and Man the Man Jesus Christ." Note that Christ is a man, and nothing else. Article 5 takes a different tack: "The Father is immoveable, no place being capable of becoming emptier or fuller of Him then it is by the eternal necessity of Nature." This in stark contrast to material quantities of which places in the world can by more or less empty, and rarefied up to diminution to infinity, as he explains to Cotes in the context of his definition of inertia. See J.E. Mcguire, "Body and Void and Newton's De Mundi Systemate: Some New Sources", *Archive for History of Exact Sciences* 3, 3 (1966): 206-248. I owe a transcript of the ms. as well as the analysis of the theological references and resonances to the generosity of Stephen Snobelen.

[31] A. Grillmeier, *Christ in Christian Tradition, vol. 1, From the Apostolic Age to Chalcedon (451)* (London, 1975).

[32] Levitin, *Ancient Wisdom*, 441.

[33] M. Frede, "Monotheism and Pagan Philosophy in Late Antiquity", in *Pagan Monotheism in Late Antiquity*, P. Athanassiadi and M. Frede (eds.), (Oxford, 1999): 41-67.

[34] See for Newton's position on this topic: Iliffe, *Priest*, 141-148.

## The intellectual context: Newton's correspondence and notes

H.A.M. snelders was, to my knowledge, the first one to attract attention to the rôle of Fatio, a young, bright swiss mathematician who personally befriended Huygens and Newton, and was in correspondence with both of them and with Leibniz for many years. This correspondence helps to understand the intellectual processes shaping Newton's ideas in the crucial period before and after the publication of the first edition of the *Principia*. The exchanges that are taking place between Huygens, Leibniz and Newton through Fatio [38] as central intermediary precedes the notorious Leibniz-Clarke correspondence by more than two decades, and is very different in tone and atmosphere.[39] Fatio at some point communicates to Leibniz Newton's reaction to Huygens's theory of gravity in an appendix *Discours de la cause de la pesanteur*, to his *Traité de la Lumière* (1690):

---

[35] See the Newton Project's catalogue entry: https://www.newtonproject.ox.ac.uk/view/texts/normalized/THEM00117 The manuscripts extant are N563M3 P222 Clark Memorial Library MS, and Keynes MS 10. Transcriptions are due to R. Iliffe, Imperial College, London, and S. Snobelen, King's College, Halifax, Canada.

[36] I analysed the *Quaestio*-structure of this manuscript in detail in an unpublished paper: K. Verelst, *On form and content in Newton's work. Epistemological dualism as an editorial strategy: a case study* (2002), https://www.academia.edu/49054782/On_form_and_content_in_Newtons_work_Epistemological_dualism_as_an_editorial_strategy_a_case_study

The topic of the relation between the Father and the Son is announced in Quest. 12, where the dispute on the word "homousios" is introduced for the first time, and reaches its summit in the Quest. [16]-[19]. These four *Questiones* occupy more than one-thirth of the totality of the Clark MS. The Trinitarian position rejected by Newton is linked to the concept of *homousia*, used in the Nicene Creed to to describe the consubstantial nature of the Father and the Son. Grillmeier, *Christian Tradition*. Also Iliffe, *Priest,* 144-148.

[37] M. Grabmann, *Die Geschichte der scholastischen Methode*, Akademische Druck- und Verlagsanstalt, 2 vols. (Graz, 1957 [1909]).

[38] For more on Fatio: S. Mandelbrote, "Fatio, Nicolas, of Duillier", *Oxford Dictionary of National Biography*, Oxford (2004): 136-140. S. Mandelbrote, "The heterodox career of Nicolas Fatio de Duillier", in J. Brooke and I. Maclean (eds.), *Heterodoxy in early modern science and religion*, Oxford (2005): 263-96. On Fatio's crucial rôle in the development of Newton's alchemy, see K. Figala and U. Petzold, "Physics and Poetry: Fatio de Duillier's *Ecloga* on Newton's *Principia*", *Archives Internationales d'Histoire des Sciences*, 37 (1987): 316-349.

[39] Cfr. the comments by Hall and Hall, *Unpublished Scientific Papers*, 205-207.

Fatio to Leibniz, April 9, 1694[40]

> Pag. 163 du Traité de Mr Hugens: Monsr. Newton est encore indeterminé entre ces deux sentiments. Le premier **que la cause de la pesanteur soit inherente dans la matière par un Loi immédiate du Créateur de l'Univers** et l'autre que la Pesanteur soit produite par la cause Mechanique que j'en ai trouvée (...)

For Fatio's claim with respect to his own contribution to the debate, I refer to Newton's contemporary draft addition to Prop. VI, Cor. 4 and 5 of the *Principia.*[41] The correspondence with Bentley, which takes place in the same period, not only confirms Newton's readiness to reconsider his opinions at the time, but also the deep connection his positions in natural philosophy held to his theological stance. It makes the "extraordinary clause"[42] in the third letter to Bentley even more extraordinary:

> Gravity must be caused by an Agent acting constantly according to certain laws; but whhether this agent be material or immaterial, I have left to the consideration of my Readers.[43]

Bentley's sharp questioning also led Newton to re-examine a number of cosmological issues in a way that went beyond the *Principia*. The relationship between God, space, and time is explicitly addressed. In a number of excerpts he refers to the Hebrew conception of the "mãqom"; God as the "Place in which we live and move and have our Being."[44] This idea will come back in the controversies with Leibniz later in his life, in fragments which Newton made as drafts for Des Maizeaux's *Recueil.*[45] For our purposes they are relevant and useful because they demonstrate again relevant interconnections between the classics and theology in the recesses of Newton's mind (see below).

---

[40] (via De Beyrie) [OH 2853; April 9, 1694]. Christtiaan Huygens, *Oeuvres complètes, Tome X. Correspondance* 1691-1695, D.J. Korteweg (ed.) (Den Haag, 1905).

[41] Hall and Hall, *Unpublished Scientific Papers*, 312-313. For Fatio's rôle in the preparation of the second edition, Cohen, *Introduction*, 177-187.

[42] Cohen, *Papers and Letters*, 276.

[43] Third Letter to Bentley, in Cohen, *Papers and Letters*, 303. Another remarkable quote from that same letter concerns the supposed contradictions present in the notion of infinity, in the mathematical sense of infinitedly small, an in the theological sense of infinitedly large. Newton plainly claims in Nature such logical inconsistencies do really exist. Cohen, o.c., 304. For a discussion of the presence of such paradoxes in both Newton and Leibniz, see K. Verelst, "Newton vs. Leibniz: Intransparency vs. Inconsistency", *Synthese*, 191, 13 (2014): 2907-2940.

[44] McGuire, "Newton on Place, Time and God", 214, 120. For the later controversy with respect to these notions, A. Koyré and I. Bernard Cohen, "Newton and the Leibniz--Clarke correspondence", *Archives Internationales d'Historie des Sciences*, 15 (1962): 94-96.

[45] Koyré and Cohen, "Newton and the Leibniz--Clarke correspondence", 94-101.

So, whatever Newton's specific position on the Trinity, it can hardly be maintained that his combined interest in professing a particularly strong postion with respect to God's fundamental unity '("Unum") (and related "titles")[46] in the General Scholium and in unpacking historical theological disputes regarding it, has nothing to do with his stance on the philosophical and theological debates ongoing in his day, however "unconvincing" this may seem.[47] The thorny issue is complicated by the fact that scholars tend to be guided in their reading of the material by the Newton they want to see rather than by the Newton they see. Morever, Newton invented and determined to a large extend himself the disciplinary requirements of 'experimental philosophy' — in its pure form, Iliffe's argument would therefore not yet be applicable to him, only to his successors.[48] The larger context to this debate is indeed the one concerning the precise status of the "Scientific Revolution" in our cultural history. That subject lies outside the scope of the present paper; I refer the reader to the interesting and relevant volume dedicated to it, edited by Margaret Osler.[49]

Be that as it may, whatever we care to think about the precise interconnection between Newton's philosophical[50] and theological preoccupations, the source material is there and it is not going anywhere. A study of the classical influences working on Newton based on his own manuscripts confirms the connection throughout his life in a matter-of-fact like way, independently of one's own *a priori* epistemological of philosophical position. The task at hand is to elucidate how precisely these different intellectual sources informed and nourished Newton at different stages of his intellectual career, how precisely he dealt with them, and what are their subsequent mutual theoretical implications. A modified 'holistic' position seems perfectly compatible with a certain level of 'disciplinary compar-tementalisation'.[51] The methods of inquiry and presentation Newton used are discipline-specific, but the same underlying fundamental epistemological and/or philosophical questions and attitude prevail throughout.[52] I will have ample occasion in this contribution to add some factual evidence to the growing pile supporting such a non-dogmatic approach.[53] The 'classical' aspect of Newton's mental universe as well as of his intellectual instrumentarium remains underestimated in much of the historiographical literature concerning him. I intend to remedy this situation somewhat, by focussing on the key text to which this volume is dedicated, the General Scholium to the second edition of Newton's *Principia*.

---

[46] De Smet and Verelst, "Platonic and Stoic Legacy". For Newton's interest in the philological and editorial aspects of the study of the Church Fathers, see Scott Mandelbrote, "'Than this Nothing can be Plainer': Isaac Newton Reads the Fathers," in G. Frank, T. Leinkauf and M. Wriedt (eds.), *Die Patristik in der Frühen Neuzeit*, Friedrich Fromm Verlag (Stuttgart, 2006): 277-297.
[47] Levitin, *Ancient Wisdom*, 434.
[48] Hence I agree with A. Janiak, *Isaac Newton. Philosophical Writings*, Cambridge (2004), Introduction, p. xii: "(...) it is obvious that the *Principia's* greater impact on the eighteenth century is to have affected a separation between technical physics on the one hand, and philosophy on the other. (...) to achieve an understanding of how Newton himself understood natural philosohy, we must carefully bracket such historical developments."
[49] M.J. Osler (ed.), *Rethinking the Scientific Revolution*, Leiden (2005), 215-42. Cambridge (2000). Also P. Barker and R. Ariew, *Revolution and Continuity. Essays in the History and Philosophy of Early Modern Science*, Washington D.C. (1991). And, of course, Dear, *Discipline and Experience.*
[50] In the two senses Newton gives to that term, so in various degrees it refers to 'science of nature', of the 'hard' kind, "experimental philosophy" or in the more 'soft', speculative sense of "natural philosophy". Cfr. Cohen, *Introduction*, 244-245. Cfr. Anstey, "Experimental versus Speculative", 215-216; 220.

## Newton and the Classics: a few general observations

Newton's dealings with ancient literature left their imprint clearly and in many ways on his own, published as well as unpublished, work. We can identify a wide array of articulation points between Newton's writings and ancient sources, ranging from direct, precisely referenced quotations collected in notebooks to much more subliminal paraphrases, footnotes and glosses, comments *in margine* of a main body of text, elucidating its content. The use of *scholia* and *glossae* as intellectual tools was, in a way similar to the *quaestio*, part and parcel of a venerable scholarly tradition, transmitted through the university curriculum of the day.[54] Newton is very well aware of their potential power[55] and uses them deliberatedly when he develops his thoughts on a certain subject in private notebooks, and even while putting them in a final form, ready for publication, like in the General Scholium. They allow him to frame in a controlled way the larger context of his thought and thus grant the reader a precious glimpse of his mental processes. Moreover, the use of ancient sources did not limit itself to his more humanistically inspired endeavours. The mathematical style of presentation in the *Principia* — which is not necessarily identical to the method of invention used[56] — is set up deliberatedly so as to satisfy to the criteria of the methods used by the Greek geometers. In a memorndum of around july 1694, in the midst of the preparations for the revised edition, David Gregory records that Newton planned two supplements to it,

> one about the geometry of the ancients where the errors of the moderns about the mind of the ancients are detected (…) [and] where it will be shown that or specious algebra is fit enough for making the discoveries, but quite unfit to [give them] literary [form] and to bequeath them to posterity.[57]

---

[51] Iliffe, "Dsciplines", 428.

[52] This evidently does not exclude the possibility that his attitude towards them evolves. Cfr. ft. 17 (Cohen).

[53] I thus take Rupert A. Hall's maxim to heart: "I dislike dichotomies of two propositions, so often neither *a* nor *b* by itself can be wholly true." Cited in J. E. Force, "The Nature of Newton's 'Holy Alliance' between Science and Religion: From the Scientific Revolution to Newton (and back again)", in *Rethinking the Scientific Revolution*, 151.

[54] M. Grabmann, *scholastischen Methode*. For Newton's educaton: S. Ducheyne, "Newton's Training in the Aristotelian Textbook Tradition: From Effects to Causes and back", *History of Science* 43 (2005): 217-237.

[55] A case in point are the *Glossa ordinaria*, the commentaries printed *in margine* of the authoritative version of the (Latin) Bible, the Vulgate. They will become the centre of an intellectual firestorm surrounding different attempts — connected to different religious obediencies — during the sixteenth and seventeenth centuries at critically editing multilingual comparative versions of the Scriptures, the so-called *polyglotta*. We will see below that Newton gets at some point cought up in this debate as well. A.E. Matter, "The Church Fathers and the *Glossa Ordinaria*". In Irena Dorota Backus (ed.). *The reception of the church fathers in the West: From the Carolingians to the Maurists*. 1. Leiden: Brill (1997): 83–111. Also S. Mandelbrote, "English Scholarship and the Greek Text of the Old Testament, 1620-1720: The Impact of Codex Alexandrinus", in A. Hessayon and N. Keene, eds., *Scripture and scholarship in early modern England*, Aldershot (Ashgate, 2006): 74-94.

The mathematical choices Newton makes thus remain firmly embedded within the orientations he opts for in his more speculative work.[58] There are also many visible traces of philologically attestable influence in the phrasing Newton uses in texts like the General Scholium without necessarily referring to an explicit source.[59] This state of affairs does not necessarily indicate a wanted omission. Many things belonged to well established common places within the cultural *milieux* to which Newton belonged — he breathed them in on a daily basis, so to say.[60] To complicate matters, we shall see that there are instances of sources Newton used intensely but then deliberatedly quenched from sight. Newton did read classical poetry and literature as well, and studied some authors (of direct philosophical interest, like Virgil or Lucretius) in detail.[61] Thus also these strands — philosophical and poetical — combine, hence it is worthwhile to follow up potential philosophical influences transmitted through poetry. A first, and promising, step in this direction has been taken by Pablo Toribio in his recent study of Lucretian ideas working on Newton through Virgil.[62] Poetry, in any case, seems to be an acceptable way for Newton to import otherwise unacceptable ideas and domesticate them into his own intellectual realm, even up to the point that they become communicable to the outside world.[63] The General Scholium proves this point in even more respects than Toribio had in mind.[64] I shall come back to this below.

---

[56] Although this is a very complex question, only partially elucidated by Whiteside's publication of Newton's mathematical papers. The classic treatment is I.B. Cohen, *Newtonian Style and Scientific Revolution*, Cambridge, 1981. For a different view, see the interesting study by De Gandt, "Le style mathématique des *Principia* de Newton", *Revue d'Histoire des sciences* 39, 3 (1986): 195-222. And of course Guicciardini, *Reading the Principia*.

[57] Cohen, *Introduction*, 193-194.

[58] Cfr. Turnbull, *Correspondence,* III, 329 (Memor. Greg. 5,6,7 May): "Appolonius' book on the determinate section of a line is in order for the cartesian problem of the ancients: for from this book one point is found, and through it and four given points the conic section is drawn. In the projected edition of these books of Appolonius at Oxford there should be a preface on the geometry of the Ancients (for which purpose the preface to the fifth or seventh book of Pappus should be consulted — it is extant in Greek as well as Arabic. Also on the conditions which make the section determinate." (transl. Turnbull)

[59] Cfr. De Smet and Verelst, "Platonic and Stoic legacy" for a philological analysis of the text of the General Scholium from this perspective.

[60] Cfr. the popularity of some Christian Neostoic writers. Lipsius's *De constantia*, was a real bestseller. J. Sellars, "Justus Lipsius's *De Constatia*: A Stoic Spiritual Exercise", *Poetics Today*, 28, 3 (2007): 341-342. Certain catch phrases got currency, cfr. paper Snobelen PNEM paper

[61] See Cohen, "'Quantum in se est", for a detailed study of Newton's dealings with Virgil and Lucretius.

There are two main periods of interest from this 'classics'-viewpoint on Newton's life: his early, formative years and the incredibly rich and complicated period in between the publication of the first and second edition of the *Principia*. For each period we happen to have a principal source. Newton's student years with respect to these influences are documented in his early notebooks, primarily the Trinity Notebook.[65] For the period of his intellectual prime there are the Classical Scholia.[66] Another line of influence runs through the early Church Fathers who documented abundantly pagan philosophy in order to refute it. They were a major scholarly resource, affecting Newton's theological writings, from where ideas and phrasings pour over again in the natural philosophical work.[67] I investigate in what follows what these sources tell us about Newton's intellectual processess at crucial steps in his career, and how, more specificially, they shed light on the genesis of the General Scholium. My focus will be mainly, though not exclusively, on material found in the Classical Scholia, and on the discrete, but intense exchanges Newton had on this material with the circle of his intimi. Thus my paper also contributes to a somewhat neglected aspect of Newton's intellectual biography: the evolution of his conceptual framework and the intellectual networks in which it evolves.[68]

---

62 Toribio, *Ita sentiebant veteres*. See also P. Toribio, XXX (this volume)

63 An interesting case in point is the dangerous, because potentially atheist, atomistic philosophy of Lucretius, which hides in plain sight in Halley's Ode at the beginning of the *Principia*. W.R. Albury, "Halley's Ode on the *Principia* of Newton and the Epicurean Revival in England", *Journal of the History of Ideas*, 39, 1 (1978): 24-43.

64 This method of using literature to make philosophical points has venerable predecessors within the Stoic tradition, with relevant authors like Cicero and Aratus. See H.C. Baird, "Stoicism in the Stars. Cicero's Aratea in De Natura Deorum", *Latomus*, 77, 3 (2018): 646-670.

65 Not only the *Quaestiones* (from 1664 onwards), but his earlier course notes as well. For the latter, one can consult the transcript on The Newton Project. McGuire and Tamny, *Trinity Notebook*.

66 Cfr. ft. 11.

67 Snobelen REFS "'The true frame of Nature': Isaac Newton, heresy and the reformation of natural philosophy". In *Heterodoxy in early modern science and religion,* John Brooke and Ian Maclean (eds.), Oxford (2005): 223-262. Mandelbrote, "Fathers". Iliffe, *Priest*: "(…) there were oceans of relevant data in the editions of patristic writings that Newton had in his own rooms and that were available to him in Trinity College Library.", 142.

68 I support Scott Mandelbrote's criticism of Dobbs's *Janus Faces* book, namely that it pays little or no attention to his contemporary context and to the intellectual communities of which he was a part — public or private. "The Janus Faces of Genius: The Role of Alchemy in Newton's Thought by Betty Jo Teeter Dobbs." Review by Scott Manderlbrote. In *The British Journal for the History of Science*, 26, 4 (1993): 491-493.

The picture that emerges before us is the following: the three editions of the *Principia* are the outward manifestation of a process of development and revision going on continuoulsy and almost entirely behind the screens — and Newton deliberatedly wanted to keep it there. Newton's work on ancient sources helps us to reconstruct this overall process, because it provides us with markers that allow us to relate it to its outward manifestations, i.e., to the different *Principia* editions, as well as to the ongoing debates with his opponents. What we have to do is to carefully place the published editions within the context of the unpublished material related to them, especially, but not only, to the projected revisions. Correspondence, discarded drafts, critical notes, shared material all contribute to the global picture. In this paper, my goal is to contribute to this picture by looking at the Classical Scholia and as much of the directly related material as possible. Interestingly, each period of Newton's involvement with the classics goes hand in hand with its own major controversy. The first one, dealing mostly with mechanical argumens, I call "early Newton". It starts during his students' days, accompagnies the period of his first discoveries in dynamics as a yong Lucasian professor and his attempts to make philosophically sense of them, in the run-up to the first edition of the *Principia*. Newton still doubts his core philosophical intuitions, and needs the encouragement from people like Halley to overcome his self-doubt in order to decide to take the step and to publisch his discoveries.[69] The controversy that marks out ths first period is the one with Hooke, a frustrating and vicious affair, which marked Newton for the rest of his life.[70] The second period, in which his involvement with classical studies and natural philosophy comes to full fruition, is during a turbulent but crucial period with heated debates and real exchange of ideas with Huygens and Leibniz, in between the first and the second edition of the *Principia*. This period I call the "middle Newton". The last period, the period of the great controversies on priority and theology, I call the "late Newton". By then the ideas are set, the debate focused on method and became dogmatic in tone. We will not deal with this last period, but point to a few interesting manuscript fragments that indicate the underlying continuity in the development of the late Newton's ideas.

---

[69] See Newton's own admittance on this in a later draft for the Preface to the second and third edition of the Principia [MS Add. 3965 f620], Hall and Hall, *Scientific Papers*, 307-308. For a detailed reconstruction of the episode, see Rigaud, *Historical Essay*, 77 sq.

[70] J. Lohne, "Hooke versus Newton. An Analysis of the documents", *Centaurus* 1, 1 (1960): 6-52. A. Rupert Hall and Richard S. Westfall, "Did Hooke concede to Newton", *Isis*, 58, 3, 4012-405. M. Nauuenberg, "Hooke's and Newton's Contributions to the Early Development of Orbital Dynamics and the Theory of Universal Gravitation", *Early Science and Medicine*, 10, 4 (2005): 518-528.

Newton's methodological approach of the classic source-texts in general can be gleaned without too much difficulty from a detailed study not only of his manuscripts, but also from his use of his books — his own[71], and those at his disposal in the collections available at Trinity and at Cambridge university Library.[72] I shall soon present some telling examples of the later period, but let us for now follow McGuire and Tamny in their description of Newton's working method while composing his *Quaestiones* as a young student.[73] They discuss in some detail Newton's efforts (in a lemma 'on motion') to reconstruct Diodorus Cronus's complex argument on time atoms. Following the traces laid out by dog-earing and annotation, they show that Newton's first-order approximation of his subject are Diogenes Laertius *De vitis dogmatis* and Walter Charleton's *Physiologia*. But Newton also owned the Paris 1621 edition of Sextus Empiricus [74], not an in first instance biographical compendium, but a serious philosophical source containing original text-material relating to pre-Socratic philosophy in which Diodorus's argument is discussed in much more detail. Charleton uses the same source, but McGuire and Tamny show that young Newton cites passages other than Charleton's to related topics, and to which no other evident access was available to him. The strong circumstantial evidence notwithstanding, McGuire and Tamny cautiously call the suggestion that Newton digged deeper into his problem by going directly to the source-text a "speculation".[75] In the section on Virgil I materially prove that this is indeed the roadmap of the procedure Newton systematically follows when dealing with this type of source-material, and very deliberatedly so. His approach is basically two-layered: a first level aims at getting an overview of basic data and the sources extant with respect to some author or topic, by reading either ancient, either contemporary, commented compilations of excerpta, translations, and biographies. He was well acquainted with

[71] As listed in Harrison's catalogue to Newton's personal library, which is to a large extend (around 60%) preserved in the collections of Wren's Library. See Harrison, *Library*, 56.

[72] "Newton's individualist epistemological self-presentation rested on the apparent right and duty of people like himself to interpret 'disputable places' in Scripture according both to implicit criteria of comprehensibility and to settled interpretive techniques. Nevertheless, in his theological research he was not working alone but was a member of a group that included John Mill, author of the major Greek New Testament of 1707 and John Covel, Master of Christ's College, Cambridge, and possessor of a number of Scriptural exemplars brought back from his stay in the Ottoman Empire. At the start of 1687, when printing of an earlier manifestation of his New Testament was well underway, Mill wrote to Covel about some of these manuscripts and asked him to present his services to 'Dr Montague, Mr Professor Newton, and Mr Laughton'. John Montagu was Master of Trinity College, Cambridge from 1683 to 1700, and John Laughton — who was almost certainly the most frequent visitor at Newton's rooms in the late 1680s - was University Librarian from 1686 to 1712." R. Iliffe, "Friendly Criticism. Richard Simon, John Locke, Isaac Newton and the Johannine Comma", in *Scripture and Scholarship in early Modern England*, 140.

[73] McGuire & Tamny, *Trinity Notebook*, 22-24, 79-89.

[74] H 1503. This copy is not preserved in the Wren Library.

[75] McGuire and Tamny, *Trinity Notebook*, 84.

Greek and read and spoke Latin fluently; several books from his student years contain annotations and translations directly from the Greek text.[76] Newton's attitude towards this material was one of genuine interest and knowledgeability. He obviously took it seriously and sought to work with it in an at once efficient and rigorous way. Although Newton, for reasons of intellectual economy, often refers at first instance to trustworthy second hand literature, he does not hesitate to go back to first rate editions of original sources, which were readily available to him in the Cambridge University Library, and of which he moreover possessed several in his personal library.[77] This close-reading of the source-text itself constitutes the second level of his approach. It should no surprise us that he followed the same method when dealing with theological mateerial.[78] The reasons for his two-step move can vary: to complete a citation; to corroborate his own interpretation, or confirm or rebuke an interpretative or linguistic position occupied by an intellectual opponent, or a combination of these. From emendations added in his own hand it will be clear that Newton's philological knowledge was as good as anyone's when it comes to critically dealing with the source-texts. Of all these we find eloquent examples in Newton's discussion of sources concerning the Church's history, especially when sensitive theological distinctions are at stake.[79] I will discuss instances of such double-layered cases stemming from both philosophical and theological contexts in what follows.

## The Classical Scholia[80]

As we saw, periods of hightened interest in ancient sources correspond with crucial transition periods in Newton's overal intellectual development. This type of material did not only influence Newton attestably while conceiving his ideas during his formative years[81], but also afterwards, culminating much later in the text that concerns us here, the General Scholium. In the period between the publication of the first and the second edition of the *Principia*, Newton arguably revisits many of his basic assumptions, putting them to the test in direct and indirect exchanges with other natural philosophers of his day.[82] Indeed, it is no exaggeration to say that the second edition reflects the fruits of this review exercise, the result of which is to largely consolidate his own basic philosophical and methodological assumptions.[83] The Classical Scholia fit in this context of revision. This becomes very clear when we include Newton's own comments to the fragments he quotes in our analysis. The subjecs discussed range from precursors to the heliocentric system over the constitution of the world soul (*spiritus*), and how it permeates and governs phenomena, to cosmology (whether the world is infinite, whether it has a center, and how the force that keeps the world together operates).[84] The Scholia consist of an annotated collection of quotes, paraphrases and extended excerpta stemming from a wide range of different sources: Presocratic, Orphic, classical, and Greek and Latin poetry, enhanced by systematic references to appropriate paragraphs from Book III of the *Principia*, collected in parallel order to lend the authority of the *Veteres*, the ancients, to his own natural philosophical work.[85] They close with a separate chapter on Macrobius's comments on Cicero's *Dream of Scipio*, of which the latter part contains a cosmology of basically Stoic inspiration.[86] Newton worked intensively with Fatio de Duiller and David Gregory to develop and mature his ideas on these topics, but finally discarded the plan to publish them in the second edition of the *Principia*, probably to avoid being drawn into the sensitive philosophical and theological controversies of his time.[87] The work was not lost on him however, because we see concepts and explicit references to it reappear in the General Scholium he added to the second edition. It is therefore highly relevant that an early version of the text of the General Scholium is part of the manuscript of the Classical Scholia, even though this remarkable fact was hardly picked up in the relevant literature, and the text is not included in any edition of them either. Let us start our examination of the material by rectifying that omission.

[76] Cfr. McGuire and Tamny, *Trinity Notebook*. I saw several examples of this in e.g, Newton's edition of Macrobius in the Wren Library.
[77] First rate even by present-day standards. Examples include: Valeriano's Virgil [H1676], Stephanus's Plutarch [H1330], Pindar [H1317], Casaubon's Diogenes Laertius [H519], Scaliger's Aristophanes [H79].
[78] "On the whole, he consulted llatin translations of original Greek writings but he often pored over Marguerin de le Bigne's *Magna Bibliotheca Veterum Patrum*, and of course, he knew the Greek New Testament extremely well." Iliffe, *Priest,* 142.
[79] Iliffe, "comma". Mandelbrote, "Fathers"
[80] Cfr. ft. 10.
[81] See for a detailed discussion McGuire and Tamny, *Trinity Notebook*, Introduction, 20-22. Also Cohen, "Quantum in se est", 139-142.

---

[82]Hall and Hall, *Unpublished*, 205-207. I.B. Cohen, *Isaac Newton's* Principia. *Introduction* (Cambridge, 1971): 172-206. Important but underestimated in this context is the Huygens-Leibniz-Fatio-Newton correspondence mentioned before (between 1686 and 1695). I presented it in my Scaliger lecture at Leiden university in 2008, and again at HOPOS 2014 in Ghent. I discuss aspects of it in my preprint on the Cornell Arxiv, K. Verelst, "when everything hinges on Gravitation", 2010, and in my Synthese paper: Verelst, "Intransparency vs. Inconsistency". I am currently preparing an edition of this correspondence which will be published by Brill. For the Arxiv paper, see https://arxiv.org/ftp/arxiv/papers/1009/1009.3053.pdf

[83] On the level of methodology, some loose threads will be wrapped up only in the Principia's third edition (1726). Cfr. Shapiro, "Newton's Experimental Philosophy".

[84] These topics accurately reflect themes that led to fierce debates in Antiquity, as evidenced by Plutarch, an author often quoted by Newton, in his refutation of the Stoics. Lucretius, a core source for Newton's Classical Scholia, plays a central role in this debate, because the cosmological passages in his poem have been interpreted since Antiquity as an Epicurean refutation of Stoic views. D.E. Hahm, *The Origins of Stoic Cosmology* (Ohio, 1977): 120-124.

[85] V. Schüller, "Newton's Scholia", 213-265. See also ft. 10.

[86] "The dream of Scipio at the end of the *De republica* develops to its most explicit level Cicero's correlation between statesmanship and Stoic ethics and cosmology." M.L. Colish, *The Stoic Tradition form Antiquity to the Early Mddle Ages I*, Brill (Leiden, 1985): 94.

[87] From the start of his career Newton got cought up in controversies inside and outside the Royal Society; with Hooke and Flamsteed inside, and the Jesuits ans the 'Cartesians' (primarily Leibniz) outside. He quickly decided that it was better for his peace of mind to stay out of them — though this proved a vain hope. Iliffe, *Priest*, 123-128; 160-166. Feingold, "Mathematicians and Naturalists", 77-102.

## Craufurd on the Classical Scholia and the General Scholium

We owe the first public announcement of the discovery of some remarkable Newtonian manuscripts dedicated to the ancients in David Gregory's estate to James Craufurd Gregory already in 1829[88], during a presentation for the Royal Society at Edinburgh. He mentions them while discussing the idea of a state of "mental derangement" which supposedly struck Newton in the period of the early 1690s,[89] as well as his ensuing presumed diminished scientific activity and his — worthy of a special mention in the context of this volume — developing interest in natural theology as a consequence of his, according to Craufurd, mental depression later in his life. Craufurd discusses in some detail what became later known as the Classical Scholia primarily as a proof for the fact that Newton had not yet lost his wits. It is worthwhile citing what he says in full, because it brings out already in a clear way the link between this newly discovered manuscript and the famous text of the General Scholium.[90] It also contains a *de facto* first publication of

---

[88] Dated by Cohen 1834; with Schüller erroneously 1832. I found it back as "Notice concerning an Autograph Manuscript by Sir ISAAC NEWTON, containing some Notes upon the Third Book of the Principia, and found among the Papers of Dr DAVID GREGORY, formerly Savilian Professor of Astronomy in the University of Oxford. By JAMES CRAUFURD GREGORY, M.D., F.R.S.E., Fellow of the Royal College of Physicians of Edinburgh", *Transactions of the Royal Society of Edinburgh*, vol. 12, issue 01, 1834, with mention: "Read March 2,1829." For the mention of Gregory, 67.

[89] Based on a lemma on Newton's discoveries in the *Biographie universelle* (1822) by Jean-Baptiste Biot, Newton's first biographer (available on the Newton Project). Biot based himself on a note by Huygens dated May, 29th, 1694, where Huygens says that a certain D. Colin has told him that Newton had been mentally down since a year and a half. Cfr. Huygens's *Oeuvres complètes* X, 616, ft. 3. This note is added in *Oeuvres* to a letter to his brother Constantyn, dated June 6th, 615-616, in which Huygens claims that the English tried to hide it "Mrs les Anglois, a ce qu'il semble, avoient tasché de cacher cet incident mais en vain." Huygens repeats this news two days later in a letter to Leibniz, *OH*, p. 618. The episode is discussed at length in Rebekah Higgitt, *Recreating Newton: Newtonian Biography and the Making of Nineteenth-Century History of Science*, Pickering & Chatto (2007). On the reality behind the claim, see M. Keynes, "Balancing Newton's Mind: his singular Behaviour and his Madness of 1692-1693", *Notes Rec. R. Soc.*, 62 (2008): 289–300.

[90] Craufurd refers to a certain M. Dutens, who, on the basis of quotes from, among others Lucretius, attempts "to prove that the general principles of motion and gravitation were known to the ancients, are precisely the same as those contained in Newton's manuscript, a considerable part of which, I find, had been long before published, nearly verbatim, in the preface to the 'Astronomiae Physicae et Geometricae Elementa' of David Gregory", and than goes on give excerpta from the manuscript. J. Craufurd, "Notice concerning an autograph manuscript by Sir Isaac Newton, containing some notes upon the third book of the *Principia* and found among the papers of Dr David Gregory, formerly Savilian Professor of Astronomy in the University of Oxford", *Transactions of the Royal Society of Edinburgh*, 12 (1834): 64–76.

large parts of three drafts of the General Scholium that are part of the Classical Scholia manuscript:[91]

Here is Craufurd's commented quotation in full:

> This account of the opinions of the ancients occupies the greater part of the manuscript; but attached to it there are three very curious paragraphs. **Two of these appear to have been the first draught of the general Scholium at the end of the edition of the Principia**,[92] published in 1713, and express the same theological opinion. It is remarkable, however, that it is only in the third edition, published in 1726, (the year before Newton died), that the substance of the second of these paragraphs is found.
>
> The first paragraph expresses nearly the same ideas as some sentences in the Scholium, commencing — "Deus summus est ens, aeternum, infinitum, absolute perfectum:" The first part of it is as follows, and the expressions appear to me still more striking and sublime than those in the Scholium itself: — "Deum esse ens summe perfectum concedunt omnes. Entis autem summe perfecti Idea est ut sit substantia una, simplex, indivisibilis, viva et vivifica, ubique semper necessario existens, summe intelligens omnia, libere volens bona, voluntate efficiens possibilia, effectibus nobilioribus similitudinem propriam quantum fieri potest communicans, omnia in se continens tanquam eorum principium et locus, omnia per praesentiam substantialem cernens et regens (sicut hominis pars cogitans sentit species rerum in cerebrum delatas, et illinc regit corpus proprium,) et cum rebus omnibus, secundum leges accuratas ut naturae totius fundamentum et causa, constanter cooperans, nisi ubi aliter agere bonum est."
>
> The second paragraph expresses precisely the same idea as the sentences of the Scholium, in the edition of 1726, beginning, "A caeca necessitate metaphysica, qua? utique eadem est semper et ubique, nulla oritur rerum variatio;" and is as follows: — "Quicquid necessario existit illud semper et ubique existit, cum eadem sit necessitatis lex in locis et temporibus universis. Et hinc omnis rerum diversitas, quae in locis et temporibus diversis reperitur; ex necessitate caeca non fuit, sed a voluntate entis necessario existentis originem duxit. Solum enim ens intelligens vi voluntatis suae, secundum intellectuales rerum ideas, propter causas finales, agendo, varietatem rerum introducere potuit. Varietas autem in corporibus maxime reperitur, et corpora quae in sensus incurrunt sunt Stellae fixae, Planetae, Com etas, Terra, et eorum partes."

---

[91] Craufurd, *Notice,* 68-71
[92] Bold added.

> The third paragraph relates to the same subject as the last paragraph of the Scholium, in which, as in his *Optics*, it is well known that Newton favours the hypothesis of a subtile and universally pervading Ether. But it is singular that it expresses upon this subject an opinion different from, and perhaps some may think sounder than, that which was afterwards published. This paragraph begins as follows: — "Coelos et spatium universum aliqui materia fluida subtilissima implent, sed cujus existentia nee sensibus patet nee ullis argumentis convincitur, sed hypotheseos alicujus gratia praecario assumitur. Quinimo si et rationi fidendum sit et sensibus, materia ilia e rerum natura exulabit;" and then proceeds to give reasons for this opinion, of the validity of which I do not pretend to judge*.[93]
>
> This manuscript bears no date, but two circumstances enable me to state that it must have been written certainly eleven, and in all probability fifteen, years before the publication of the second edition of the *Principia* in 1713. *1st,* The edition of David Gregory's *Elements of Astronomy*, into which, as already stated, much of that portion of the manuscript which relates to the opinions of the ancients has been transferred almost verbatim, was published in the year 1702. *2d,* I find the whole of the manuscript fairly copied, in the handwriting of David Gregory, into the end of a manuscript book, containing his unpublished notes upon the *Principia* of Newton, and bearing a running date from 1687 to 1697.

Craufurd read his "Notice concerning an autograph manuscript by Sir Isaac Newton" for the Royal Society of Edinburg on the second of March, 1829.[94] The autograph manuscript to which Craufurd refers is today's Royal Society MS 247. This is the most significant version of what is now known as Newton's Classical Scholia, a series of additions in his own hand to parts of Book III of the *Principia* he considered for inclusion in its re-edition in the 1690s.[95] These notes or scholia are part of Newton's effort to show that his natural philosophy was already embraced in whole or part by ancient Greek and Roman natural philosophers. In 1838, Rigaud refers to Craufurd's discovery in 1829 in his detailed account of the gestation and publication history of the first edition of Newton's *Principia*, quoting the latter's presentation:[96]

[93] [Footnote*] The remainder of this paragraph is as follows: "Nam quomodo motus in pleno peragatur intelligi non potest; cum partes materiae, utcunque minutae, si globulares sint, nunquam implebunt spatium solidum; sin angulares, propter omnimodum superficierum contactum firmius haerebunt inter se quam lapides in acervo, et ordine semel turbato, non amplius congruent ad spatium solidum implendum. Porro tam experimentis probavimus quam rationibus mathematicis, quod corpus sphaericum densitatis cujuscunque in fluido ejusdem densitatis utcunque subtili progrediens, ex resistentia medii prius amittet semissem motus sui quam longitudinem diametri suae descripserit. Et quod resistentia fluidi illius nee per subtilem partium divisionem, nee per motum partium inter se diminui possit, ut corpus longitudinem diametri prius describat quam amittat semissem motus."

[94] Craufurd, "Notice", 64–76.

[95] Another, sligthly different, copy in Gregory's hand, is available as part of Gregory's *Notae* on Newton's *Principia*, RS 210. On the relation between the extant versions Gregory's *Notae* and the Classical Scholia, see W.P.D. Wightman, "David Gregory's Commentary on Newton's 'Principia'", *Nature*, 4556, February 23 (1957): 393 – 394. Cfr. ft. 10.

> But he only mentions that he 'found (along with several other autograph fragments on mathematical subjects) one maniscript consisting of twelve folio pages in the handwriting of Newton, and containing, in the form of additions and scholia to some propositions in the third book of the Principia, an account of the opinions of the ancient philosophers on gravitation and motion, and on natural theology, with various quotations from his works'.

Afterwards the Classical Scholia manuscript with its addendum disappears again from sight for more than a century, only to reappear in a footnote to Turnbull's *Correspondence* edition of David Gregory's 1694 Memoranda of his visit to Newton.[97] In a next step, Hall and Hall publish excerpts of related material on Lucretius and Aristotle, together with a one-page facsimile reproduction.[98] Cohen, in a well-known article on Lucretius's influence on Descartes's and Newton's formulation of the inertial law,[99] links Gregory's Memorandum to the excerpt published by the Halls, and explains that not only Lucretius's, but as well Aristotle's argument of continued infinite motion in the absence of any resistance (*Physics* IV, 8) was construed by Newton as a precursor to his own First Law:

> Most historians of science today would hesitate to attribute to Aristotle any true share in formulating the Law of Inertia. It is therefore all the more interesting that Newton himself once included Aristotle among those ancient thinkers who knew the first law. Newton quoted two extracts from Aristotle to support his contention: the one (from *Physica* IV,8) we have printed above, and another from *De caelo* 3, 2.[100]

Cohen dedicates a paragraph to a manuscript given by Newton to Gregory, and containing exerpta from Roman and Greek authors. Another important point is the connection to Richard Bentley. The noted classicist had been working on Lucretius in view of a new edition in the same period he was exchanging letters with Newton on the core tenets of the latter's natural philosophy and their theological implications, in preparation of his Boyle Lectures in 1692.[101] Given Bentley's status in philology, one could reasonably assume that Newton discussed his findings in this field with him in the period they worked together on the second edition of the *Principia*. Such conversations could account for the condensed and cleaned out version of the Classical Scholia that Newton at some point prepared for the second edition, and inserted in one of his annotated and interfleafed copies of the 1687 edition.[102] This version is not included in any existing Classical Scholia edition, and I do not find it discussed anywhere. It is relevant, however, because of the selection it takes from the ancient source-material in the lager versions, and also because of the entirely different citation strategy it follows when compared to the published General Scholium.

[96] S.P. Rigaud, *Historical Essay*, 80.
[97] Turnbull, *Correspondence* 3, 336-339, especially ft. 10.
[98] Hall and Hall, *Unpublished Scientific Papers*, Plate V, 304, 309-311.
[99] Cohen, "Quantum in se est", 139-146.
[100] Cohen, *Quantum is se est*, 141.

Towards the end of his *Quantum*-paper, Cohen announces his intention to publish the Classical Scholia material, but this publication never materialises. McGuire and Rattansi give the first thorough, but contested,[103] interpretation of the manuscript in their notorious "Pipes of Pan" article in 1966. B.J.T. Dobbs follows suit in her groundbreaking book on Newton's *Alchemy* of 1975. For a full publication of the Classical Scholia, however, we have to wait until 1981 when Casini brings out his Italian paper, and then its translation into English in 1984.[104] Some emendations and corrections to this edition are proposed in De Smet and Verelst.[105] A recent edition by Schüller, first as a Working Paper at the Max Planck Institut (in German), and then in an English translation, concludes provisionally the fateful editorial history of this remarkable document.[106] I say "provisionally" because, although excellent, Schüller's edition cannot be the end of the story, its claim about itself notwithstanding, because he does not take all the extant material into account and still contains several mistakes.[107]

[101] Cohen, "Quantum in se est", 148–149. Cohen already published the four letters Newton wrote to Bentley (as well as part of the latter's two sermons on "A Confutation of Atheism") in his collection of Newton material: I. Bernard Cohen, *Isaac Newton's papers and Letters on Natural Philosophy* (Cambridge, 1958): 271-394. Cfr. Turnbull, *Correspondence* 3, 313, 317.

[102] On sheets inserted in variant $E_1i$ of the *variorum*-edition. Bentley was close enough to the revision process in the early years of 1700 that we find annotations and corrections in his hand on two other of Newton's personal copies of the printed edition which was being prepared for revision. What makes me suspicious about potential Bentleyan influence, apart from the differences in citation strategy, is the fact that it is precisely everything referring to Lucretius in the original version of the Classical Scholia has dissappeared. For details about the different *Principia* copies being prepared and Bentley's potential rôle in some of them, see the Apparatus criticus of the *variorum*-edition: A. Koyré and I. Bernard Cohen (with A. Whitman), *Isaac Newton's Philosophiae Naturalis Principia Mathematica, The Third Edition with Variant readings*, vol. 1 (Cambridge, 1972): .ix-xii. For the $E_1i$- version of CS: *id.*, vol. 2, 803-807. See also Cohen, *Introducion*, 189

[103] Notoriously , "The Classical Scholia". But also Ducheyne, "General Scholium", ft. 9.

[104] Casini, "The Classical Scholia".

[105] De Smet and Verelst, "Platnic and Stoic Legacy".

[106] Schüller, "Newton's Scholia". For more details, see ft. 10.

## The Classical Scholia's RS 247 missing page (folio 14r)

Craufurd astutely identifies the three paragraphs of folio 14r of the Classica Scholia as containing arguments later found in the General Scholium.[108] He moreover correctly notes that the arguments in the second paragraph do not appear in print until the 1726 edition of the General Scholium appeared. As we saw, Craufurd supplies substantial transcriptions from these three paragraphs. In their celebrated 1966 paper on the Classical Scholia, J. E. McGuire and P. M. Rattansi do note f14r and conclude: "It is plain therefore that the theological ideas in the General Scholium had been sketched by the early 1690's, and that Newton saw these being compatible with the philosophy presented in the scholia to the key propositions of Book III".[109] In his 1984 study of the Classical Scholia, Paolo Casini does not mention the three paragraphs on Royal Society MS 247, f14r. Although he does include folio 14 when he speaks of the range of folios in the version of the Classical Scholia found in MS 247, in his transcription from this manuscript he completely omits folio 14r. He does include in his 1984 paper a photoreprint of the English edition (London, 1715 and 1726) of David Gregory's publication of material from the Classical Scholia that originally appeared in his *Elementa astronomiae physicae et geometricae* (London, 1702).[110] Gregory's 1702 publication also omits the material from Royal Society MS 247, f14r. Volker Schüller does not mention folio 14r, either. In sum, there has been an almost total lack of attention paid to f14r. However, its importance should be clear not only from its early date, but also because it demonstrates directly textual and conceptual relationships between the Classical Scholia of the 1690s and the General Scholium of the 1710s and 1720s. Let us look closer at these two points.

Codicological analysis[111] of the manuscript preserved at the Royal Society in London gives the following results: the text on folio 14r is scratched out by a thick line running diagonally over it. The last paragraph is scratched out additionally by a large cross. This page is present in the original Newton manuscript contained in RS MS 247, but not in

---

[107] "in einer hoffentlich endgültigen form edieren," Schüller, 1. But e.g. the summary of the CS-material in the interleaved copy $E_1$i (cfr. *variorum*, Appendix III, 803-807) is not accounted for.
[108] Cfr. ft. 88.
[109] McGuire and Rattansi, "Pipes of Pan". The quote is from 139, n. 3.
[110] Casini, "The Classical Scholia", 47–58.
[111] During a visit of the Royal Society's Library in London, in 2018.

Gregory's neat copy RS MS 210. The handwriting and material outlook give no reason to think that this page is extraneous to the manuscript, or a later addition. However, folio 14r appears to be unique in that it does not contain editorial notes about placement in the *Principia*.[112] This along with the fact that it is crossed out may also have been the main criterion of exclusion used by McGuire and Rattansi, Casini and Schüller. Not that it makes the editorial decision any more acceptable.

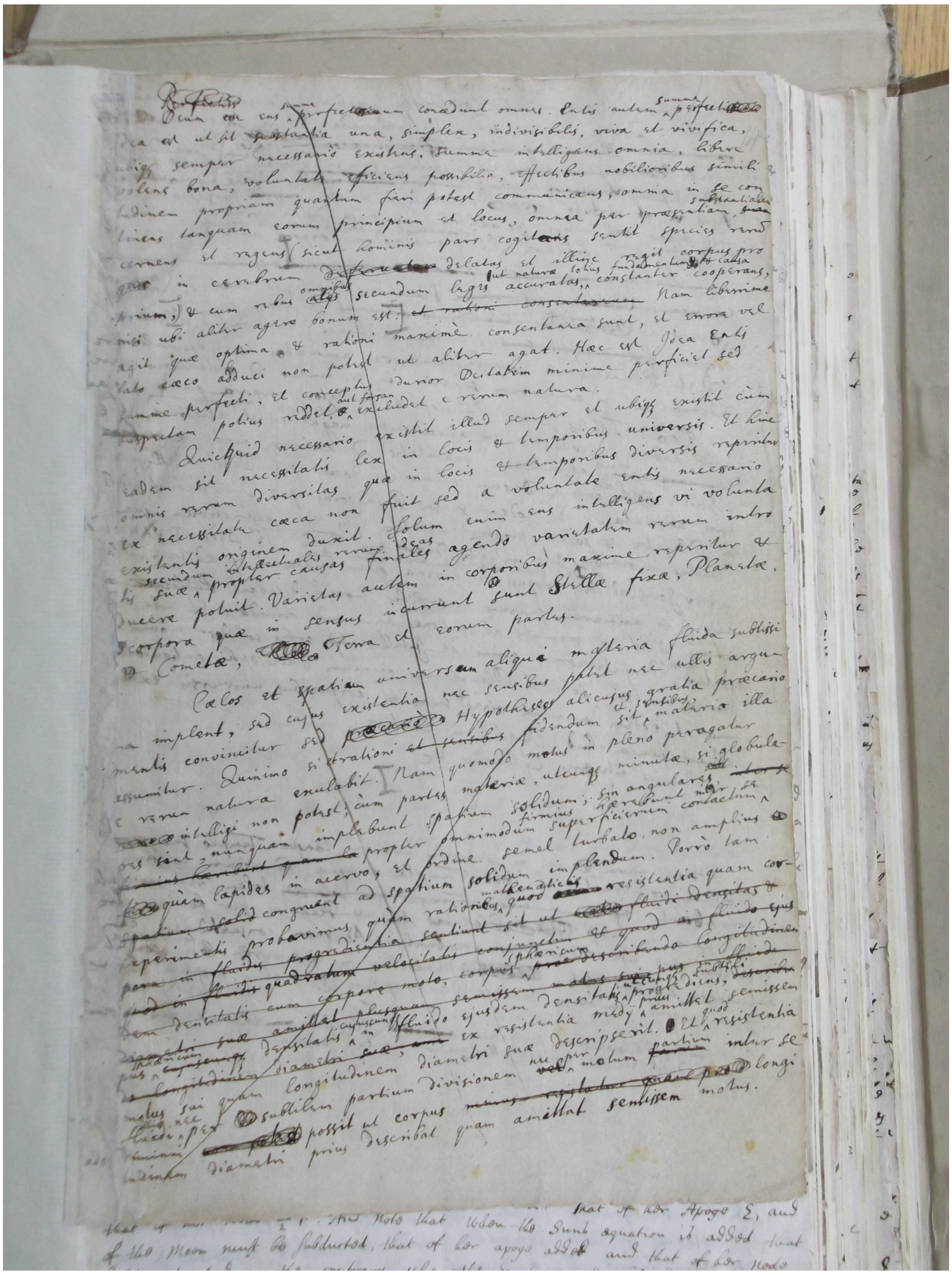

*The f14r "missing page" from RS MS 247*[113]

[112] I owe this observation to a personal communication by Steffen Ducheyne.
[113]

On the basis of pictures taken, we can see that there are some watermarks which might provide indications on the date of the mss. The watermark on RS 247 folio 10r (which we find back again on f12r[114]) resembles closely Gravell watermark SLD.317.1, dated November 4, 1686. The watermark on RS 247 f14r (our 'missing page'[115]) is the nearly same as the Gravell watermark SLD.204.1, dated November 1686. They are in all likelihood variants from the same papermill and the same supplier during rougly the same period, which might run through several years.[116]

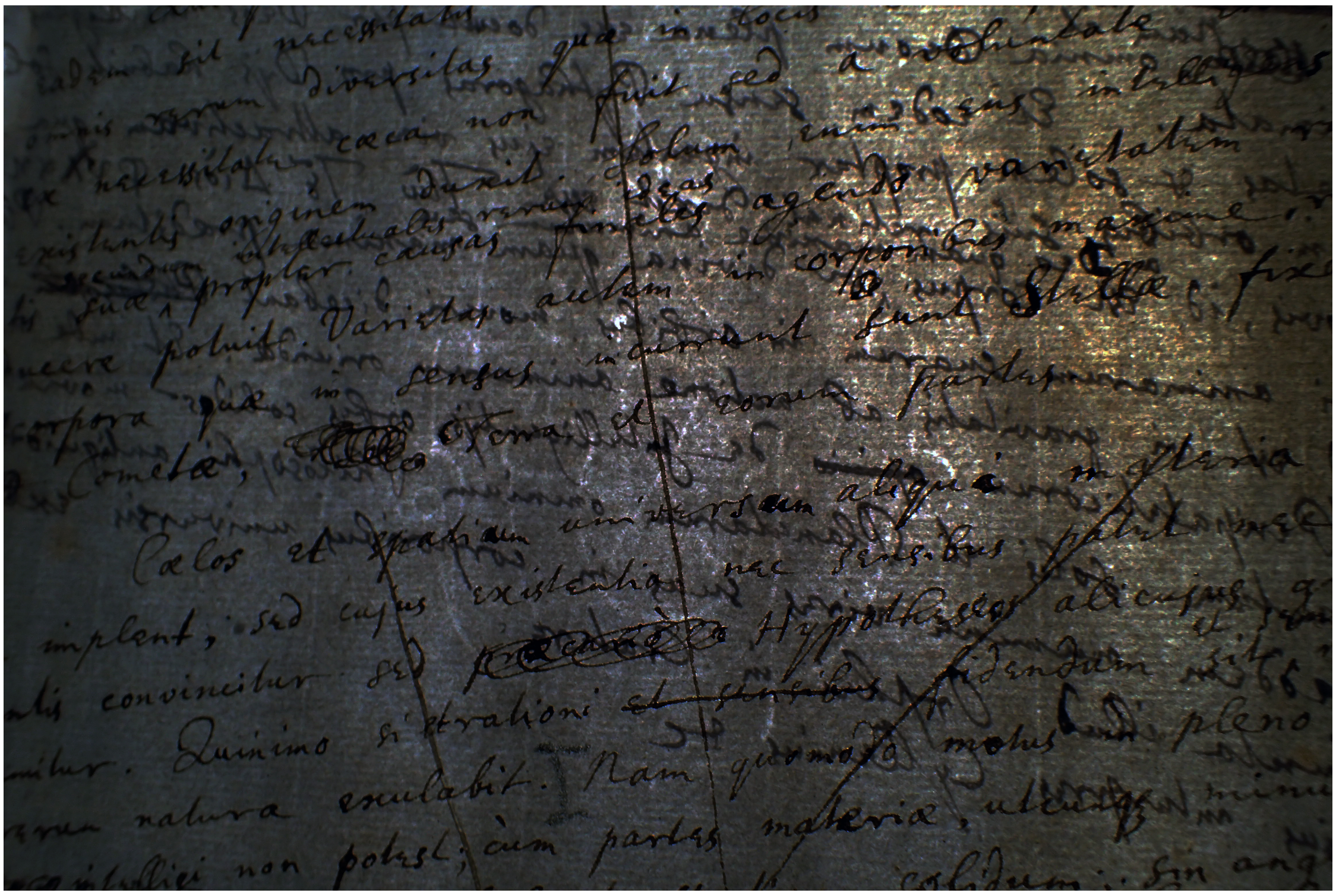

*Watermark on RS MS 247 f14r*

---

[114] A shield with a fleur de lys.

[115] An ornament with a hanging horn.

[116] On the tedious art of establishing chronlogical links between manuscripts by means of watermarks: Alan E. Shapiro, "Beyond the dating game: watermark clusters and the composition of Newton's *Opticks*", in: *The investigation of difficult things. Essays on Newton and the history of the exact sciences in honour of D. T. Whiteside*, P.M. Harman and Alan E. Shapiro (eds.), Cambridge (1992): 181-229. Of course, by "the same", I mean under the limited possibilities of scrutiny at our present disposal: for RS 247, by means of photoshop enhanced versions on screen and on print (Courtesy Stephen Snobelen) vs. the rather low quality images (on screen and in print) that I found on the Gravell website. But still, the fact that we get such a 'hit' referring to a common material origin on at least two different places strongly supports the case for the unity of the CS collection, more precisely, that f14r is not a later addition or somehow slipped into it by error. It would be nice to be able to compare the size and the material of the paper used in both cases with that of RS 247 to see if it is equal. Or to find out whether there are different sizes and kinds of paper commonly produced with similar watermarks over the same time period by a given producer, and accessible both in London and Cambridge.

Now, what could have been the motivation for Newton to set out on this painstaking enterprise of collecting supporting fragments throughout ancient philosophical literature and organise them in a systematic way, in view of incorporation into the second edition of the P*rincipia*? Newton moreover did not shy away from sharing his progress in the form of private notes to some of his trusted friends, and to use his findings as (in his view) solid arguments in informal discussions on his opponents' points of view. This was much more than just a rhetorical device in an intellectual dispute. This is clear from a text copy made by his amanuensis, Humphrey Newton, of a manuscript *De motu corporum liber secundus*, which was originally planned to become Book II (the later Book III) of the *Principia*.[117] The text was written in an accessible, non-mathematical, "philosophical" way and presented the implications of his discoveries in dynamics for a description of "the System of the World" — a title under which it would be published posthumously both in Latin and in English in 1728. According to Rigaud, he changed his mind about publishing it because of the ongoing controversy with Hooke on the priority of universal gravitation and the inverse square law.[118] In the first paragraphs, Newton goes at some length to explain that "the Romans, the Egyptians, the Greeks and the Chaldeans" taught a heliocentric worldview in which the planets moved about the sun through empty space. He moreover mentions, interestingly, that we do not know how the ancients explained the forces that kept the planets in their course.[119] For many years Newton entertained the idea to re-edit the *Principia* while incorporating these ancient philosophical allies, a plan he finally discarded.[120] The work was not lost on him, however, as traces of his endeavour continued to survive implicitly and explicitly in, among others, the text that concerns us here, the General Scholium to the second edition of the *Principia*.

[117] Newton, *System. Introduction*, ix.

[118] See the detailed (and revealing) reconstruction by Rigaud, *History of Principia*, 63-67. A modern, more contextual and sociological point of view (doing more justice to Hooke) in N. Guicciardini, "The Hooke-Newton Debate on Gravitation: Recent Results", *Early Science and Medicine*, 10, 5 (2005): 510-517. Interestingly, Newton was not considered to be the experimentalist in the first phase of the debate, after Newton's presentation of his paper on light for the Royal Society in february 1672. When the controversy reignited in 1685, Newton decided to erase the name "Hooke" from the text of the *Principia*, as he states in a letter to Halley in 1886. See M. Nauenberg, "Hooke's and Newton's Contributtions to the Early Development of Orbital Dynamics and the theory of Universal Gravitation", *Early Science and Medicine*, 10, 4 (2005): 518-528. On the internal debates and tensions within the Royal Society with respect to the status of "experimentation" vs. "speculation" and the attitude towards the Aristotelian philosophical tradition, the background against which all this plays out, see Ducheyne, *Main Business*, especially Ch. 2, and Levitin, "Robert Boyle". An interesting approach to the debate from a philosophy of language point of view — the growing role of mathematical language in claims toward authority in the nascent community of experimental philosophers, is S. Shapin, "Robert Boyle and Mathematics: Reality, Representation, and Experimental Practice", *Science in Context* 2 (1988): 23-58.

[119] Newton, *System*, 4.

[120] As he makes clear in a letter to Cotes, 2 March, 1712/13. Hall and Hall, *Unpublished Scientific Papers*, 186 (ft 1).

> Newton was genuinly anxious to discuss his theory of matter in detail and publicly. Three times he tried to find a place for it in the *Principia*: in the *Conclusio*, in the Preface and in the General Scholium; and three times he rejected it. (...) But all that at length emerged after painful reflection was a cautious hint in the printed version of the Preface to the first edition, to be followed years later by the oracular but confusing conclusion to the General Scholium to the second edition. No one can be sure of knowing the reasons for the ultimate suppression of his cherished ideas. (...) the *Principia* was the work which Newton was the most anxious to make immune from attack. His second thoughts were no doubt tactically wise, and helped to avoid bitter controversy (...)[121]

Buchwald and Feingold seem to think along similar lines, since they suggest that Newton was triggered to publish them even more by the publication of Leibniz's *Tentamen de motuum* in 1689.[122] The suggestion is interesting, but it is not clear on what base it is made. Surely, the fact that Leibniz opens the *Tentamen* with the statement that "the ancients (...) did not yet understand the splendour of nature" cannot suffice, least of all because they leave out the very relevant clause, "especially those who followed the beliefs of Aristotle and Ptolemy".[123] To understand this as a "thinly veiled jibe at the burgeoning cult of Newton" is far-fetched to say the least. Moreover, a bit further in the *Tentamen*, Leibniz calls himself upon the authority of the ancients, more particularly Leucippus and Epicurus, to support his reformulation of Cartesian vortex-theory.[124] The authors miss the opportunity to clarify their interpretation of this episode using the Leibniz-Newton correspondence, to which they nonetheless refer.[125] Their reading of the one letter from Leibniz to Newton they cite is particularly uncharitable. Contrary to what they claim, nothing indicates that Leibniz has at the moment of writing any "aggressive" intentions towards Newton. Not only does he heap praise on Newton's mathematical discoveries (calling his own contribution "an attempt that didn't end badly either"); he also calls Newton's theory of universal gravity an "amazing discovery", and adds that he is "still inclined to believe that all of these [motions] are caused or regulated by the motion of a fluid medium," immediately adding that this "would not at all detract from the value and truth of your discovery" — implying simply that he accepts Newton's mathematical results but continues to disagree with him on their physical interpretation, a perfectly rational position to take.[126] To conclude, Leibniz mentions Huygens and refers to his tract *Discours de la cause de la pesanteur*, published together with his *Traité de la lumière* in 1690, where a similar physical interpretation is defended, and asks Newton respectfully to contemplate it and to give his opinion, "for it is by the friendly collaboration of you eminent specialists in this field that the truth can best be unearthed."[127]

---

[121] Hall and Hall, *Unpublished Scientific Papers*, 186-187.
[122] "The publication of Leibniz's *Tentamen de motuum coelestium causis* in the February 1689 issue of the Acta eruditorum fortified Newton's resolve to establish the ancient origins of the true frame of nature." Buchwald and Feingold, *Civilisation*, 144.
[123] Bertoloni Meli, *Equivalence and Priority,* 126-127.
[124] Bertoloni Meli, *Equivalence and Priority*, 128. E.J. Aiton, "Leibniz on Motion in a Resisting Medium", *Archive for History of Exact Sciences*, 9, 3 (1972): 257-274.
[125] Buchwald and Feingold, *Civilisation*, 144-145.

In order to understand what is going on, the context of this letter is really important. Leibniz had published his three essays on dynamics, among which his *Tentamen,* beginning of 1689, after the publication of the *Principia.* Huygens had published his work on dynamics in 1690, after a visit to Newton and Fatio at the Royal Society in London, again after the publication of the *Principia.*[128] No one would think of accusing Huygens of aggression towards Newton for this reason, and no one should. The four of them engaged in an exchange of ideas through correspondence that lasted several years until Huygens's untimely dead in 1695, and in which Fatio, who knew all of them well,[129] played the rôle of intermediary.[130] There were many open questions in the new philosophy, and even Newton, certain as he was of the superiority of his universal mathematical description of the phenomena in nature, did not simply take for granted his personal ideas on its physical interpretation, especially on the cause of gravity, but was still prepared to reconsider his own positions in view of other researchers' arguments. I pointed out already before that this is confirmed by Newton's correspondence with Bentley. To paraphrase Cohen: Newton was  not yet a Newtonian.[131]

Leibniz's criticism with respect to using the ancients is but a weak echo of Huygens's, who in a letter to Fatio sharply rejects Newton's tendency to equate his ideas to that of philosophers of two millenia ago. Fatio had seen a version of the Classical Scholia and communicated their content to Huygens. Huygens mockingly complains about Newton's habit of referring to the ancients to support his own view as if that 'support' added anything to the debate (Huygens's own view clearly was that one can use the ancients in the same way one can use the Bible: to defend whatever position one chooses to defend). So it makes perfect sense to situate the gestation of the Classical Scholia in the period of precisely that debate. Huygens's ironical reaction is remarkable enough:

> Fatio to Huygens (15 Feb. 1692) [OH 2739; *Correspondence* 3, 383]:
>
> Monsieur Newton croit avoir decouvert assez clairement que les Anciens comme Pythagore, Platon &c. avoient toutes les demonstrations qu'il donne du veritable systeme du Monde, et qui sont fondées sur la Pesanteur qui diminue reciproquement comme les quarrez des distances augmentent. Ils faisoient dit il un grand mystere de leurs connaissances. Mais il nous reste des divers fragments, par ou il paroit, à ce qu'il pretend, si on les met ensemble, qu'effectivement ils avoient les mêmes idées qui sont repandues dans les *Principia Philosophiae Mathematica*. **Quand Monsieur Newton se seroit trompé, il marque toujours beaucoup de candeur de faire un aveu comme celui la.**[132]
>
> Huygens to Fatio (29 Feb. 1692) [*Correspondence* 3, 385[133]]:
>
> Monsieur Newton fait bien de l'honneur aux Pythagoriciens de croire qu'ils aient esté assez bons geometres pour trouver de pareilles demonstrations a celles qu'il a données touchant les Orbes Elliptiques de Planetes. Pour moy j'ay de peine à croire qu'ils avaient seulement connu le mouvement de Mars, Jupiter, et Saturne autour du Soleil, et la proportion de leurs cercles; parce que Platon ayant achevé les Ecrits de Philolaus, y auroit trouvé tout le Systeme Copernicien s'il y eust esté; et ne s'en seroit pas teu. Mais quand à la vertu centrifuge qui contrebalance la Pesanteur, j'en remarquay ces jours passez quelque vestige dans Plutarque au Traité *de facie in Orbe lunae*, ou il dit que la pesanteur de la Lune ne la fait pas descendre vers la Terre, parce que cette pesanteur est effacée par la force de son mouvement circulaire, semblable à celle qu'on sent quand on fait tourner une pierre dans une fronde.

---

126 This is also exactly Huygens's point of view, as is clear from a note dated 14 dec. 1688. C. Huygens, *Oeuvres Complètes*, vol. 21, 143. For a discussion, H.A.M. Snelders, "Christiaan Huygens and Newton's Theory of Gravitation", *Notes and records of the Royal Society of London* 43, 2 (1989): 209-222.
127 Turnbull, *Correspondence* 3, 258. The translation is Turnbull's.
128 Snelders, "Christiaan Huygens and Newton's Theory".
129 Newton asked Fatio to send a copy of his *Principia* to Huygens and Leibniz in july 1687. Huygens did the same with his *Traité de la lumière*, asking Fatio to send it to seven prominent English natural philosophers and experimenters, "many of you're your good friends, "Mrs. Newton, Boyle, Hamden, Halley, Locke et Flamsteed, lesquels vous estant tous connus, et la pluspart vos bons amis" [February 7 1690 OH 2558].
130 Hall and Hall, *Unpublished Scientific Papers*, 312-313. Cohen, *Introduction*, 177-187. Mandelbrote, *Fatio.*
131 Cohen, *Paper and Letters*, 277. See on this also Dobbs, *Janus Faces*, 194.

Buchwald's and Feingold's reading therefore fits more the tone and atmosphere of the later controversy[134] that ultimatedly led to the Leibniz-Clarke correspondence, and which was guided behind the screens by Newton.[135] By then, his attitude has indeed changed completely.[136] In a manuscript from the period 1710-1720, Newton says: [137]

[132] Turnbull, *Correspondence* 3, 193. Bold added. Huygens nevertheless admits to have "found over the last days some relic [of these ideas on gravity] in Plutarch's Treatise de facie in Orbe lunae". The in any case intriguing remark by Fatio ("when Mr. Newton is mistaken, he likes to make a whole game out of such comments") derserves more attention, but that is outside of the scope of this contribution. I will discuss the background and implications of this in my planned edition of the HLFN-correspondence.

[133] This letter as such is not included in Huygens's *Oeuvres complètes*, only some preliminary notes. It is reproduced by Gagnebin. Turnbull, *Correspondence* 3, 197.

> In the latter part of his Postscript he (...) falls foul upon my Philosophy as if I (and by consequence the ancient Phenicians & Greeks) introduced Miracles & occult qualities.[138]

"He" of course is Leibniz. In my opinion, this is one of the most stunning statements ever written by Newton. According to Koyré and Cohen, it is part of a draft for a letter to Conti (in the same period as the Leibniz/Clarke correspondence).[139] It comes after the publication of the 1713 General Scholium, and in the midst of the Leibniz-Clarke controversy. As Koyré and Cohen point out, the correspondence Newton-Conti-Leibniz reproduces the Leibniz/Clarke-debate closely. From our perspective this is important because Newton, twenty years after the composition of the Classical Scholia, reveals how essential the claim to the authority of ancient predecessors continues to be for him, and how he extracts philosophical authority in the disputes with his opponents from this presumed connection.[140]

In the Classical Scholia, Newton extensively used both popular and scholarly editions of compilations from Plutarch, Diogenes Laertius *lives of the philosophers*, sources on Democritus, Lucretius *De rerum natura*, Aristotle on Pythagoras and the atomists, Macrobius on Pythagoras, Aeschylus, Cicero, Virgil and so on. Newton's aim is not philological accuracy, but real comprehension — to which end philological accuracy may be a useful tool. He works freely with his source material according to his own needs and never hesitates to edit or rearrange it when it does not comply sufficiently to his interpretation.[141] A telling example is the manner in which he edits out in an almost chirurgical manner all explicitly Stoic material: although it deals with a crucial philosophical concept — the *spiritus* or world soul —, the Stoics are to be avoided at all costs, because the materialist conceptions they entertained of this fundamental power did not fit in Newton's natural philosoghy, nor is his theology. We will take a closer look at a particular and telling example from the Classical Scholia, in order to get a better feeling of Newton's at the same time rigorous and creative method in his dealings with the material the *Veteres* had left him, and of the goals he seeks to achieve.

[134] A. Koré and I. Bernard Cohen, "The Case of the Missing Tanquam", *Isis*, 52, 1961, 555-566.
[135] Koyré and Cohen, "Newton and the Leibniz--Clarke correspondence", 63-126; 67.
[136] Shapiro , "Newton's Experimental Philosophy", proposes a plausibe explanation for this change.
[137] Portsmouth Add. 3968, fol. 591 and 589.
http://www.newtonproject.ox.ac.uk/search/results?n=25&sr=26&ce=0&keyword=Conti&all=1&sort=id&order=asc
[138] Koyré-Cohen, "Leibniz--Clarke correspondence", 111. For a related fragment, see 110.
[139]The Newton Project classifies it as a draft for the Commercium Epistolicum.
[140] This principle extends to authority claims in other cultural contexts, like the Jewish, as well. The mention of the MAKOM (like the reference to the "phenicians" in the Opticks, query XXVIII) is the result of the belief — or wish to believe — in the antiquity, and thus the "respectability", of certain major conceptions." Koyré-Cohen, "Leibniz--Clarke correspondence", 100. For the "MAKOM", see also ft. 41.
[141] Dobbs, *Janus Faces*, 194.

## Virgil: A poetical vehicle for dangerous thoughts

It may seem a bit surprising at first glance, but Newton was an avid reader of ancient literature and poetry. His personal library included many items of which we do not find any trace in his published or unpublished work, but which bear clear marks of being consulted, allowing us to presume that even the great man read sometimes for leisure only. Nevertheless, even poetry could fit into the grander scheme of things, as is plain from the fact that two classic poets appear in the enigmatic gloss that pops up in the latter, theological part of the General Scholium: Virgil and Aratus. The 1713 version of the General Scholium [142] contains a gloss to *ipso* in the sentence:

> In ipso continentur & moventur universa, sed sine mutua passione.[143] (In him are all things contained and moved; yet neither affects the other)[144]

Newton then gives several references to classical sources, cited as authorities (*veteres*) to support its content:

> Ita sentiebant veteres, ut Pythagoras apud Ciceronem, de Natura deorum, lib. I. Thales, Anaxagoras, Virgilius, Georgic., lib. iv. v 220 & Aneid. lib. 6. v. 721. Philo allegor. lib. 1. sub initio. Aratus in Phaenom., sub initio. Ita etiam scriptores sacri ut Paulus in Act. xvii., v. 27, 28 (...)

We pointed out in earlier work that this list of references matches nearly perfectly the one given in Lipsius's *Physiologia stoicorum* and that this is no coincidence: Newton almost certainly used Lipsius's work as a guide to useful classical, but theologically defused concepts about the universal "spiritus" that somehow animates the world.[145] The remainder of the references concern different books of the Bible, to which we will come back later. The sentence itself belongs to the 'theological part' of the General Scholium. It establishes the link between God and absolute space. Toribio has shown in a nice article that Newton uses Virgil here as a vehicle to carry philosophical messages of a potentially disruptive nature, namely, the philosophy of Lucretius and its implications for Newton's notion of absolute space. He remarks that Newton, although clearly leaning on contemporary authors for his information on the classics, might have researched the sources as well.

[142] *variorum*-edition, 762. The translation I use is Motte-Cajori's.
[143] Cfr. McLaurin's comment.
[144] Transl. Motte-Cajori, *The Principia*, Prometheus, Armherst (1995): 441.
[145] De Smet and Verelst, "Platonic and Stoic Legacy", 15-16. See also Dobbs, "Stoicism". For an edition of Lipisus, see J. Lagrée, *Juste Lipse et la restauration du stoïcisme: Étude et traduction des traités stoiciens De la constance,*

> (...) the addition of those two references by Newton suggests that he personally researched the sources. In the case of Virgil, we can presume knowledge of his poems in anyone who had received a scholarly education in the seventeenth century; furthermore, Newton owned four editions of Virgil's poems, one of them being Pierio Valeriano's 1532 edition, which included Servius' *Commentarii*. And Lucretius is referred to twice by Servius in his commentary on the aforementioned *Georgics* passage.[146]

I can confirm that suggestion in the affirmative. Newton did not only consult Macrobius on Virgil, but added lines from his 1532 Virgil-edition to his excerpta not present in his secondary source, and thus in all likelihood has consulted also Servius's *commentarii* referring to Lucretius. Details follow below.

Newton privately calls Lucretius's philosophy "old and pure",[147] but he clearly is, because of his (according to Newton erroneous) affiliation to atheism, a fellow traveller too dangerous to be seen with too often in public. This concealment is part of a more general editorial strategy Newton used while endlessly redrafting the text of the General Scholium, and which Snobelen has dubbed "the principle of obliquity".[148] It allowed him to keep his heterodox inspirations behind the curtain while handing out some cues to the informed reader. The passage in Virgil's *Georgica* had already in Antiquity been related to Lucretius's ideas on the impossibility of material destruction, but is recast by Virgil to the

---

*Manuel de philosophie stoicienne, Physique des Stoiciens (extraits)* (Paris, 1994). For Lipsius's ideas on this topic, see H. Hirai, "Lipsius on the world-soul between Roman cosmic theology and Renaissance "prisca theologia"", in *Justus Lipsius and Natural Philosophy*, eds. H. Hirai and J. Papy (Brussels, 2011).

146 Toribio, *Ut sententiebant veteres*.

147 Memoranda Gregory 1694, Turnbull, *Correspondence* 3.

148 S. Snobelen, "'To discourse of God': Isaac Newton's heterodox theology and his natural philosophy", in P. Wood (ed.), *Science and dissent in England, 1688-1945*, Aldershot (2004): 39-66; also —, "Isaac Newton, heretic: the strategies of a Nicodemite", *British Journal for the History of Science* 32, 4 (1999): 381-419.

idea of the impossibility of the death of souls.[149] Interestingly, one of the four (!) editions of Virgil that Newton owned,[150] Valeriano's 1532 edition, contains Servius's *Commentarii*, in which this interpretation can be found.[151] I discussed before that Cohen pointed out already decades ago in his *Quantum*-paper that Newton understood Lucretius and Aristotle as precursors to his ideas on inertial motion in empty space, and quoted them as such in the Classical Scholia. [152] We shall see below that the concealment-technique Snobelen identified in the General Scholium is pushed through much further in the Classical Scholia which, the reader should keep this in mind, originally were intended for publication in the second edition of the *Principia*. This is relevant also because Virgil establishes one of the few outspoken links between the Classical Scholia and the General Scholium: he is quoted extensively in the former, as well as cited *in margine* in all printed versions and many drafts of the latter.[153]

Virgil, however, carries more hidden messages than just this one. At some places in the Classical Scholia this becomes very clear by simply analysing Newton's working method on specific text-instances. Newton was working systematically with both a compendium or a commentary and the first-hand, actual text sources on his writing desk. An intriguing case to illustrate the situation is offered by some fragments of the Classical Scholia out of Macrobius' *Dream of Scipio*. These fragments stand a bit separate from the actual scholia to propositions IV-IX.[154] The topic at hand is the rôle played by celestial music as the immaterial and mathematically describable force that holds the world together — and thus a conceptual precursor to Newton's own gravitational force, on the same account as we saw before for Lucretius and Aristotle for the First Law.[155] Newton cites Macrobius:

> Nullus sapientum animam ex sympnonijs <quoque> musicis constitisse dubitavit (None of the wise men doubted that the soul was <also> constituted by musical concords).
>
> Macrob. Som. Scip. 1.1.c.6.

and then sets out to discuss Cicero's idea of the sensible world as a temple: even when visible, it is the expression of something higher that can only mentally be grasped and is worthy of veneration.[156] He goes on to say that Virgil held a similar point of view, in that he attributed a soul to the world which feeds the heavens, the earth and the sees from the inside — an exhortation relevant for some passages in the General Scholium. What follows is a literal quote from Macrobius,[157]

> Hunc rerum ordinem <et> Vergilius expressi, nam et mundo animam dedit et ut puritati <eius> attestatur, mentem vocavit: caelum enim, ait, *et terras et maria & sydera spiritus intus alit*,[158] id est anima, sicut alibi pro spiramento animam dicit
>
> Quantum ignes animaeque valent
>
> & ut illius mundanae animae assereret dignitatem, mentem esse testatur:
>
> Mens agitat molem.
>
> necnon, ut ostenderet ex ipsa anima constare et animari universa quae vivunt, addidit:
>
> Inde hominum pecudumque genus.
>
> Macrob. ib. [Som. Scip.] 1. 1. c. 14.

[149] In his *commetarii*, Library, quotes Lucretius (1, 671): "continuo hoc mors est illius quod fuit ant*e*". Interesting with respect to this is a statement Gregory ascribes to Newton in his 5,6,7 May 1694 Memoranda: "Ad Religionem non requitur Status animae separatus sed resurrectio cum memoria continuata (Not a separate existence of the soul, but a resurrection with a continuation of memory is the requirement of religion)." *Correspondence* 3, 336, 339.

[150] Harrison, *Library*, 256. Only two of them remain in the Wren Library, the 1532 Valeriano edition, and the edition by the French Jesuit Charles de la Rue (C. Ruaeus) of 1696 (H 1679).

[151] H 1676. Cfr. Toribio (who refers to Harrison's page number, not to the number of the item).

[152] Cohen, "Quantum in se est", See on this also Hall and Hall, *Unpublished Scientific Papers*, 309: "Although he refers to Pappus in the first line of the Preface to the *Principia*, Newton did not often make histroical allusions to the ideas of the ancients in his published scientific writings. It is clear that this was not from ignorance, since he was well read in the classical authorities. This passage is of special interest as indicating that Newton was prepared to find antecedents for the First Law of Motion not merely in the Moderns, Galileo and Descartes, but in the ancients, Lucretius and Aristotle — an historical impulse which he later overcame." It will be clear by now that I do not subscribe to that latter statement...

[153] Cfr. Snobelen's overview of biblical references to the General Scholium.

[154] They are on ff. 7r-v. Schüller, 242-245.

---

[155] McGuire and Rattansi, "Pipes of Pan", refer to Conduitt on Newton's beliefs concerning Pythagoras in this context: "I thought Pythagoras's music of the spheres was intended to typify gravity & as he makes the sounds & notes to depend on the size of the strings, so gravity depends on the density of matter", 117.

[156] Dobbs connects this to Newton's reflections on the 'First Temple' tradition and iconography, through Augustine's *De musica*, and Macrobius, Chalcidius and Boethius. *Janus Faces*, 154-155. Augustine's *Opera* are Harrison's lemma H101 in Newton's library, but not preserved. On the 'corruption' of this "ur-religion" and the importance of the history of religon for Newton's natural philosophy, see Levitin, *Ancient Wsdom*, 436-439.

[157] From Newton's copy as preserved in the Wren Library, [H1013] Macrobius, Ambrosius Aurelius Theodosius. *Opera. Ioh Isacius Pontanus secundò recensuit: adiectis ad libros singulos notis. Quibus accedunt I. Meursii breviores notæ.* 8º, Lugduni Batavorum, 1628. p. 58 [Tr/NQ.8.70]. There is a big dog-ear from the left top corner to the sentence "Mens agitat molem" in the middle of the page.

[158] The italics are in the original.

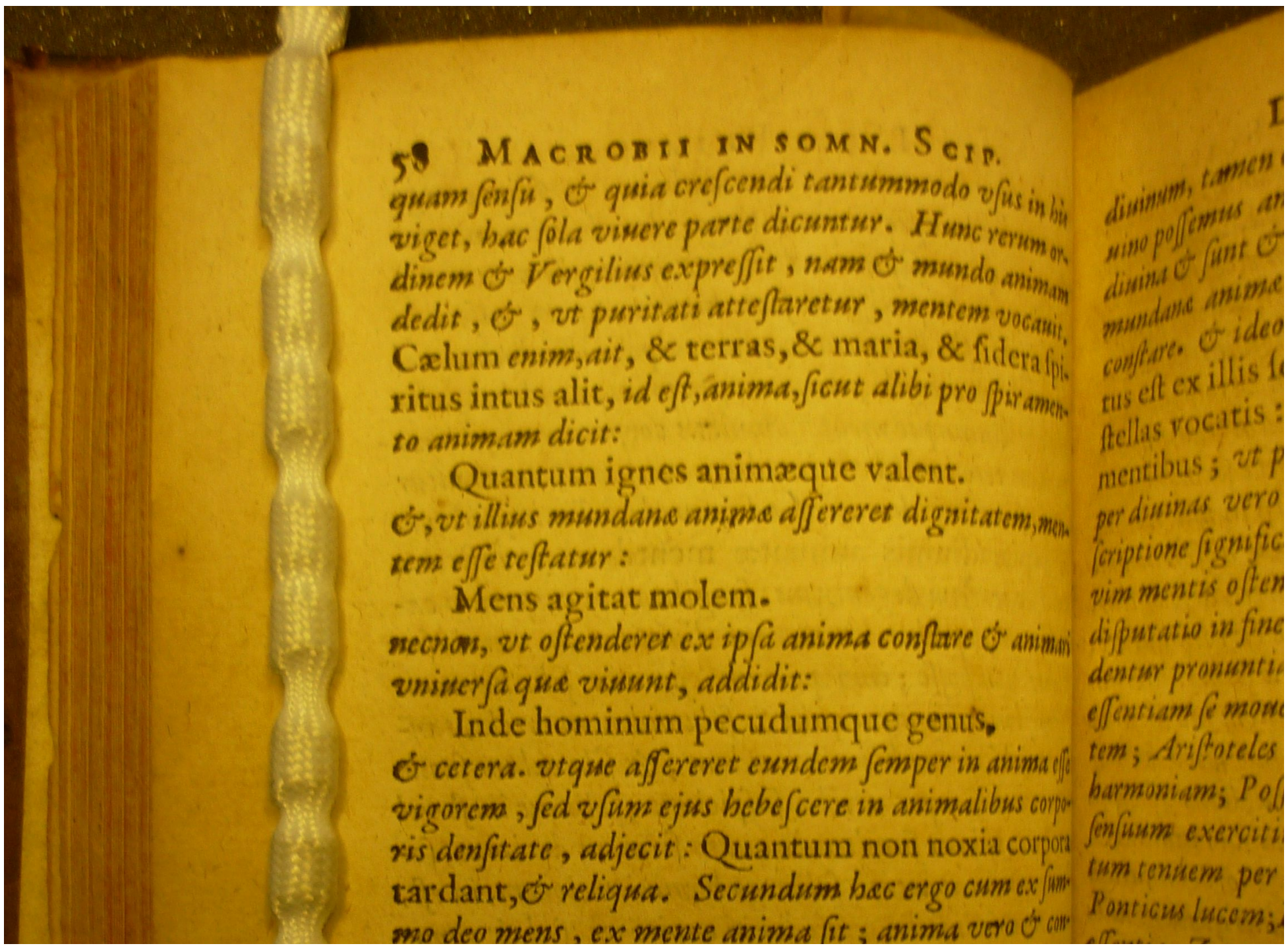

58 MACROBII IN SOMN. SCIP.
quam sensu, & quia crescendi tantummodo usus in his viget, hac sola viuere parte dicuntur. Hunc rerum ordinem & Vergilius expressit, nam & mundo animam dedit, &, ut puritati attestaretur, mentem vocauit, Cælum enim, ait, & terras, & maria, & sidera spiritus intus alit, id est, anima, sicut alibi pro spiramento animam dicit:
Quantum ignes animæque valent.
&, ut illius mundanæ animæ assereret dignitatem, mentem esse testatur:
Mens agitat molem.
necnon, ut ostenderet ex ipsa anima constare & animari vniuersa quæ viuunt, addidit:
Inde hominum pecudumque genus,
& cetera. utque assereret eundem semper in anima esse vigorem, sed usum ejus hebescere in animalibus corporis densitate, adjecit: Quantum non noxia corpora tardant, & reliqua. Secundum hæc ergo cum ex summo deo mens, ex mente anima sit; anima vero & con-

*Page from Newton's copy of Macrobius with Virgil's verse.*

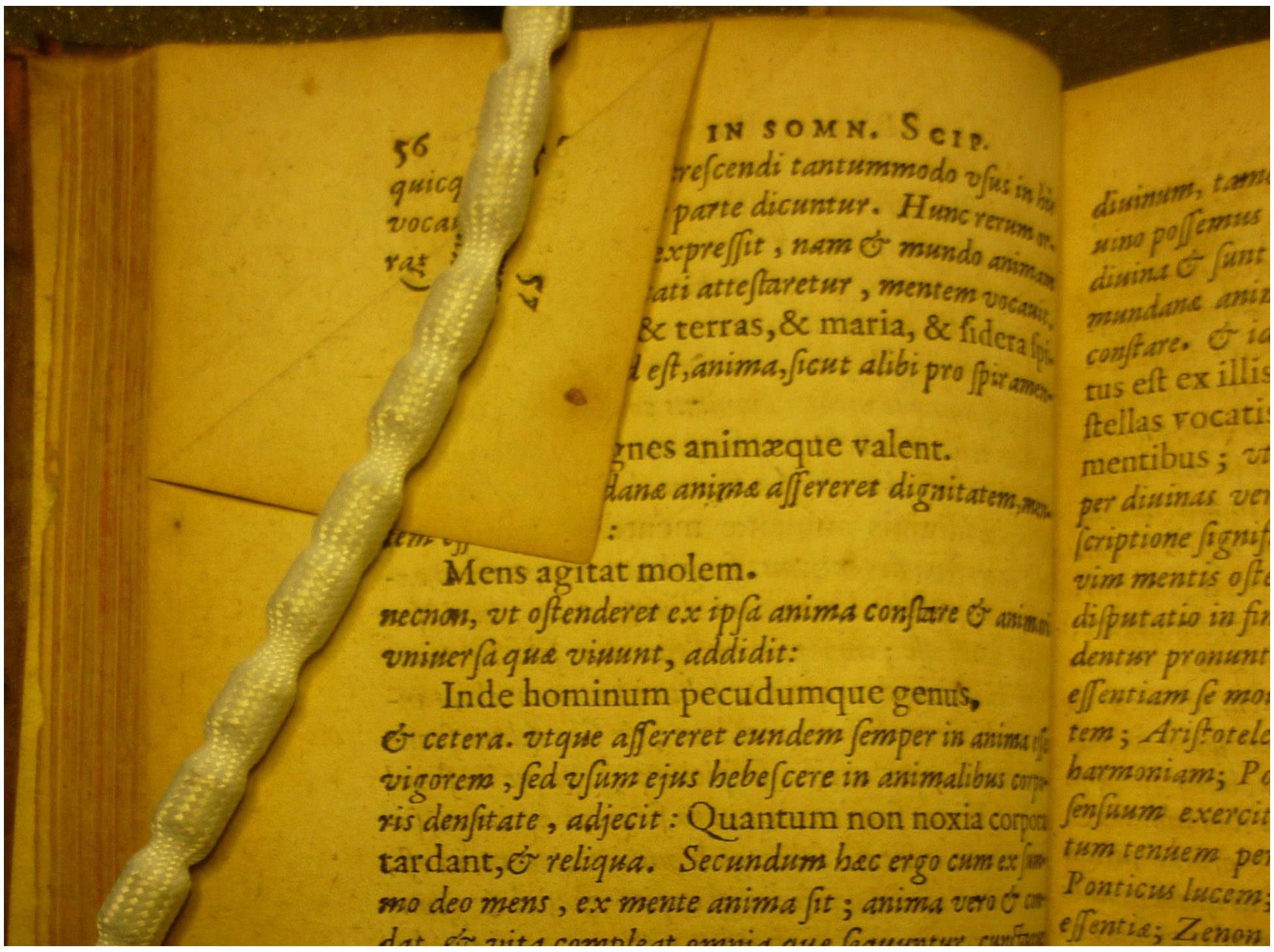

*The same page, with the relevant passage dog-eared.*[159]

But the last sentence does not end where Newton marks it with an endpoint. In Macrobius's text we find *& cetera*, indicating that in the original on which Macrobius based himself the text of Virgil continued. Macrobius then dwells at some length on the celestial fires we call *sideras et stellas*, and their relation to the fire of the eternal spirit. Newton goes on to quote from p. 59, where opinions of many ancient authors on this topic are listed,

> Plato dixit animam essentiam se moventem. Xenocrates numerum se moventem, Pythagoras et Philolaus Harmoniam, - Democritus spiritum insertum atomis, hac facilitate motus, ut corpus illi omne sit pervium
>
> Macrob. ib. 1. 1. c. 14.

[159] © Wren Library, Cambridge

Interestingly enough, this quote is incomplete in a far from innocent way. The discrete indent is the only mark that indicates that a large part of the sentence was cut out. Schüller points out in his edition that Newton left out a reference to Aristotle's *entelecheia*.[160] Disappeared as well between Philolaus and Democritus have Posidonius's *idea*, Hippocrates's subtle spirit[161], Heraclitus and Zeno's fiery versions of the material spirit, all of which are discussed by Lipsius in his *Physiologia stoicorum* as defenders of the "fiery spirit" (*spiritus igneus*)[162], i.e., a version of the world soul that is too overtly material to fit in smoothly in Newton's own explanatory scheme.

Then he jumps forward in Macrobius's text to a passage that strengthens the argument for "music" als the binding force in the cosmos according to the Pythagoreic tradition, followed by Plato:

> Hinc Plato postquam et Pythagoricae sucessione doctrinae et ingenij proprij divina profunditate <cognovit> nullam esse posse sine his numeris [sc. musicis] jugabilem competantiam: in Timaeo suo mundi animam per istorum numerorum contextionem ineffabili providentia Dei fabricatoris instituit. - - Et paulo post: Ergo mundi anima quae ad motum hoc quod videmus universitatis corpus impellit, contexta numeris musicam de se ~~XXX~~ creantibus concinentiam, necesse est ut sonos musicos de motu, quem proprio impulso praestat efficiat, quorum originem in fabrica suae contextionis invenit.
>
> Macrob. Som. Scip. 1. 2. cap. 2.

---

160 Schüller, "Newton's Scholia", fts 99 and 100.

161 The Hippocratic corpus was a major influence on the early Stoa in the development of concepts like of *pneuma* (Let. *spiritus*) in a psychophysiological sense. E. Frixione, "*Pneuma* — fire Interactions in Hippocratic Physiology", *Journal of the History of Medicine and Allied Sciences* 68, 4 (2012): 505-528.

162 Hirai, "world soul", 67.

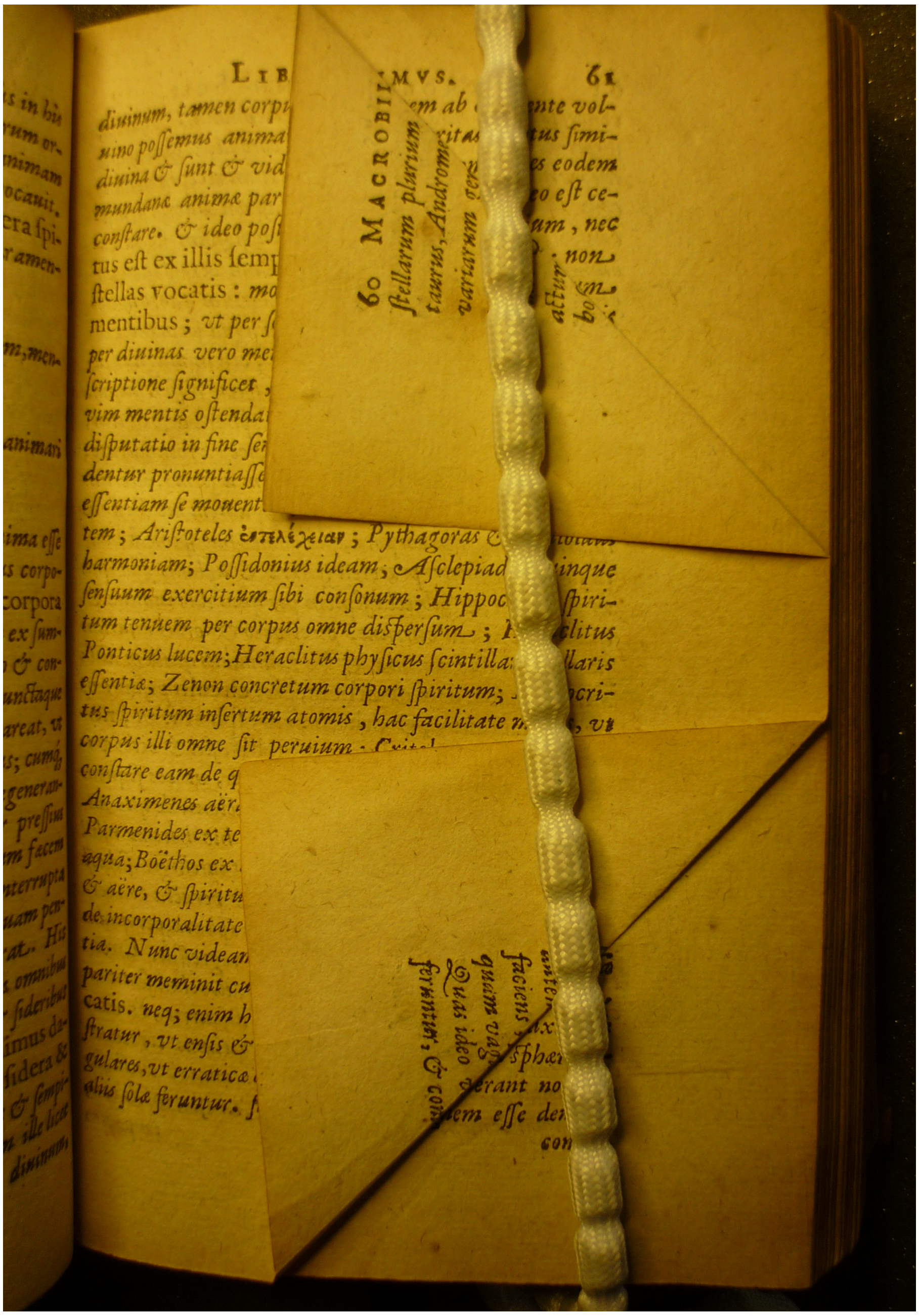

*The rejected textfragment in Macrobius fits in nicely between two dog-ears.*[163]

Instructive again, because in this case the omission is clearly accounted for ("Et paulo post"). It is on p. 104, and marked with a dog-ear. The selection here seems to be guided merely by criteria of textual economy. The last part of his dealing with Virgil in this Macrobian context gives,

> Inesse <enim> mundanae animae causa musicae quibus est intexta praediximus. Ipsa autem mundi anima ~~vitam~~ viventibus omnibus vitam ~~miXX~~ ministrat,

---

163 © Wren Library, Cambridge

hinc hominum pecudumque genus vitaeque volantum
et quae marmoreo fert monstra sub aequore pontus[164]

There is no specific reference to Macrobius for this part of the fragment, and I could nowhere find it back in Macrobius's text. The first verse evidently is the one referred to at the beginning of this excursus, but now completed; of the second verse there is no trace in Macrobius. But Newton had an excellent source for it, and we know that he used it because it is extensively dog-eared; his state-of-the-art edition of Virgil's complete works.[165]

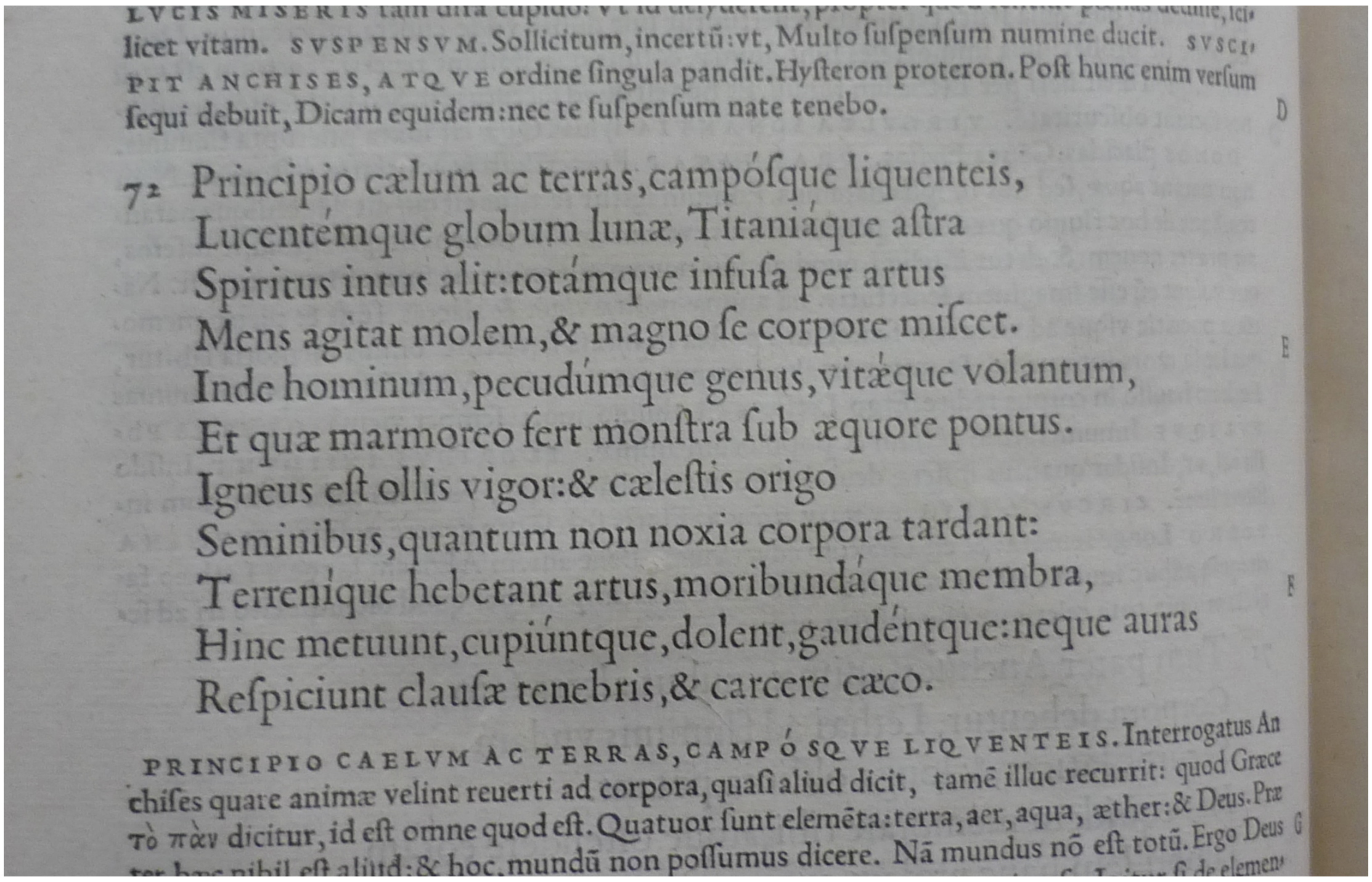
licet vitam. SVSPENSVM. Sollicitum, incertũ: vt, Multo ſuſpenſum numine ducit. SVSCIPIT ANCHISES, ATQVE ordine ſingula pandit. Hyſteron proteron. Poſt hunc enim verſum ſequi debuit, Dicam equidem: nec te ſuſpenſum nate tenebo.

72 Principio cælum ac terras, campóſque liquenteis,
Lucentémque globum lunæ, Titaniáque aſtra
Spiritus intus alit: totámque infuſa per artus
Mens agitat molem, & magno ſe corpore miſcet.
Inde hominum, pecudúmque genus, vitǽque volantum,
Et quæ marmoreo fert monſtra ſub æquore pontus.
Igneus eſt ollis vigor: & cæleſtis origo
Seminibus, quantum non noxia corpora tardant:
Terreníque hebetant artus, moribundáque membra,
Hinc metuunt, cupiúntque, dolent, gaudéntque: neque auras
Reſpiciunt clauſæ tenebris, & carcere cæco.

PRINCIPIO CAELVM AC TERRAS, CAMPÓSQVE LIQVENTEIS. Interrogatus Anchiſes quare animæ velint reuerti ad corpora, quaſi aliud dicit, tamẽ illuc recurrit: quod Græce τὸ πᾶν dicitur, id eſt omne quod eſt. Quatuor ſunt elemẽta: terra, aer, aqua, æther: & Deus. Præter hæc nihil eſt aliud: & hoc mundũ non poſſumus dicere. Nã mundus nõ eſt totũ. Ergo Deus

*Newton's 1532 Valeriano edition of Virgil, Aen. vi. 721-731, referenced in the General Scholium*[166]

The addition in itself does not seem of much consequence at first sight, but it is clear that Newton, unsatisfied with the point at which Macrobius choose to interrupt Virgil, went on to look in his own edition for what should follow. However, Newton himself also leaves out a small, but far from innocent piece of Virgil's text:

> mens agitat molem *et magno se corpore miscet* [167]
>
> (Virg., *Aen.* vi, 727)

[164] Transl. Schüller:
We strongly emphasized that the causes of harmony are traced to the world-soul with which it is interwoven, but the world-soul itself grants life to everything that is alive.
Thence come the race of man and beast, and the life of the birds
And what creatures the ocean hides beneath its marble mirror

[165] It is not clear why Newton has 'hinc' instead of 'inde' at the start. Both Macrobius and Valeriano have 'inde'. P. Vergilius Maro, *Opera. Mauri Seruii Honorati grammatici in eadem commentarii ... Castigationes & varietates Virgilianae lectionis, per I. Pierium Valerianum.* Fo, Parisiis, 1532. The final quote stems from Aen. vi, 723-6, exactly as Newton indicates in his gloss (except that he refers to the passage starting from the first line).

[166]

In literal translation this gives: "the mind stirs the mass and mixes itself with the great body". The first part we found back already in the quote from Macrobius. The same verse, accompagnied by the preceding sentence "spiritus intus alit, totamque infusa per artus", is quoted and cut off by Lipsius at exactly the same place, interestlingly enough in a passage in which he critically discusses Stoic ideas on the relationship between God and the world soul that pervades everything. He refers to Cicero's discussion of Pythagoras's version of the *anima mundi*, and quotes him verbatim: "the world is wise and has a mens that constructed itself and the world, and **orders, moves and governs all**".[168]

Newton's philosophical cleanup job could hardly become more obvious. His fragment ends where also Virgil ends the sentence in the stanza.[169] Upon closer inspection one realises that the stanza starts with the verse from the Aeneid referenced in the General Scholium. One should note that the fact that the last verse which connects the beings in the seas[170] to the binding effect of the celestial 'music' elsewhere resonates with the explicit connection between phenomena in the heavens and the seas (the tides) under the influence of gravity in the General Scholium. Once we put it side by side with two other

---

[167] Italics added

[168] Hirai, "World-Soul", 67-68. The fragment stems from Cicero, *Academica*, II, xxxvvii, 119, a passage to which also Newton refers (bold added). Cfr. Schüller, "Newton's Scholia", 230. Hirai adds: "This passage is so important to the Flemish philosopher that he quotes it on several occasions throughout his treatise". It is clear that the concordances between Lipsius and Newton are not just "loose textual parallels hat do not convince", as Levitin claims. *Ancient Wisdom*, 434, ft. 545.

[169] The translation of the complete stanza:
"[Aniches replies] First, the heaven and earth, and the watery
plains, the shining orb of the moon and Titan's star,
a spirit within sustains, and mind, pervading its
:members, sways the whole mass and mingles with its
:mighty frame. Thence the race of man and beast,
the life of winged things, and the strange shapes
ocean bears beneath his glassy floor. Fiery is the
vigour and divine the source of those life-seeds, so far
as harmful bodies clog them not, nor earthly limbs
and mortal frames dull them. Hence their fears and
desires, their griefs and joys; nor discern the the
light, pent up in the gloom of their dark dungeon."
Cfr. *Virgil in Two volumes*, H. Rushton Fairclough and G. P. Goold (eds.), Loeb Classical Library 63, Cambridge, MA, 1916, 556.

[170] The reader may be reminded of prop. XXIV. Theorema XIX of Book III, on the tides..

fragments, the scholia to props. VIII and IX, the addition obtains its full meaning: music acts, akin to gravity, as an immaterial force whose effects can be expressed by mathematical laws[171]:

To Prop. VIII[172]

> The ratio with which gravity decreases as the inner distance from the planet increases was not sufficiently explained by the ancients. The appear to have concealed this ratio using the harmony of the celestial spheres, whereby they portrayed the sun and the remaining six planets (...) with the seven-strynged lyre and measured the intervals between the spheres through the tone intervals. (...) Therefore Macrobius says (Book I Chap. 19) The lyre of Apollo with seven strings, as many as there are celestial spheres, enables us to understand their movements, for which nature has appointed the sun as ruler. (...) Therefore they called the god Hebdomageta, the sovereign of the number seven. [Λ a. Ἑπτ. ἐπὶ Θηβ. v. 739. Aeschylus: ὁ σεμνὸς ἑβδομαγετής Ἄναζ Ἀπόλλων. Venerable Hebdomageta., King Apollo] (...)

To Prop. IX[173]
Schol.

> Up to this point I have explained the properties of gravity. I have not made the slightest consideration about its cause. However, I would like to relate what the ancients thought about this. Thales believed every body to be animate and concluded this from the magnetic and electric attractions. (...) Pythagoras believed the sun and the planets to be gods and said that the sun, because of its powerful force of attraction, is τὴν τοῦ Διὸς φυλακήν Jupiter's jail, that is, a body endowed with the greatest possible divine force, with which the planets are locked inside their shells. For the mystic philosophers, Pan was the highest divine creature, who inflamed this world and melodically played it like a muscial instrument with the harmonic ratio, in accordance with that <verse> by Orpheus Ἁϱμονίαν κόσμοιο κϱέκων φιλοπαίγμονι μολπῇ. Therefore they called both God and harmony the souls of the world composed by harmonic numbers. (...)

---

171 McGuire and Rattansi, "Pipes of Pan", 117. Dobbs, *Janus Faces*, 154-155.

172 Transl. Schüller, 235. Prop. VIII: "If two globes gravitate toward each other, and their matter is homogeneous on all sides in regions that are equally distant from their centers, then the weight of either globe toward the other will be inversely as the square of the distance between the center." *id.*, 216.

173 Transl. Schüller, *Newton's Scholia*, Prop. IX, 239.

The last fragment gives us once more a nice opportunity to see Newton the scholar at work. Apart from Conti's compendium on mythology cited in Schüller, Newton consulted his excellent edition of the Orphic hymns[174], where the verse can be found on dog-eared on p. 108. I do not know whether this accounts for the different reading Schüller gives in his comment (μολπᾳ, which he reads[175] in stead of μολπῇ), because Newton's copy of Conti's *Mythologiae* is not preserved in the Wren Library. Newton's rendering is in agreement with the Stephanus-Scaliger edition of the Orphica, however.

To sum up: Newton intervenes in Macrobius's text to complete the Virgilian stanza with a verse on the connection between the world soul that imparts life and movement to things and the celestial music that originates from these motions, and grants through its musical harmonic intervals the order and structure of the universe, exactly as does his own universal force of gravity. He takes however utmost care to avoid anything that would support a materialist interpretation of this world soul. Nevertheless, this reference to a structuring divine mind that pervades everything leads us in a natural manner to ancient Stoicism.

## The Elephant in the Room: Stoicism

The impact of Stoicicm on Newton's thought is contested. In her pioneering work, Dobbs demonstrated that Stoicism played an important rôle in the development of Newton's alchemical ideas, and through them in the development of his natural philosophy.[176] But there has been controversy surrounding her work, claiming hat she overestimated Stoicisms' impact to the detriment of other important influences working on Newton, such as Neoplatonism.[177] Or, on the other hand, that such influences are overrated in any case, since Newton freed himself already early on from such "metaphysical" considerations.[178] I do not wish to take a position in this debate, nor do I claim that Newton is a Stoic, a Neoplatonist, or the opposite.[179] The only thing that

174 A commented critical edition by Stephanus and Scaliger, published in 1689 — thus a book he might have acquired in the period he was working on the revision of his *Principia.*
175 Schüller, *Newton's Scholia*,, ft. 76.
176 Dobbs, "Newton and Stoicism"; —, *Janus Faces.*
177 McGuire and Rattansi, "Pipes of Pan". Casini, "The Classical Scholia"
178 H. Stein, "Newton's metaphysics", in *The Cambridge Companion to Newton*, I.B. Cohen and G.E. Smith (eds.) (Cambridge 2002): 256-307.
179 I do agree, however, with Madenlbrote's criticism of Dobbs's work, tha it neglects the inellectual and communities contemporary to Newton when developing his ideas. Manderbrote, *Review*, 492.

interests me are Newton's ancient sources and what we can learn from his use of them. What we learn is that Newton has a vivid interest in many topics dealt with by the Stoics as well as their contemporary opponents, especially with respect to the relation between God and space and the operation of the universal "spirit" in our bodies and in the world. This concern visibly permeates parts of the General Scholium. We also see that Newton carefully selects and manages his sources, and discusses the implications of the different versions of this *spiritus* with respect to its relation with bodies, the void and the center of the world. This line of research fits in nicely with the research programme Newton laid out decades earlier in the *De gravitatione*: if we could discover experimentally how we move our bodies, we might learn something on how God moves bodies from one part of space to the other.[180]

The relationship betweeen the mind and the body, and between man and God, would become a fundamental element of Newton's theological metaphysics. (…) Significantly, it cnstituted a major and explicit atack on those sections of Descartes's Principia that dealt with notions of rest and moement, the nature of extension, the relationship between mind and body, and the relationship between God and extension.

Clearly, there is a connection between Newton's philosophical and his theological concerns, as we saw already in the section on Virgil. And indeed, Newton was an avid student of theology, Church history, chronology, and the interpretation of prophecy, on all of which he wrote elaborate notebooks[181] and treatises[182], some of them published soon after his death.[183] To this end, he undertook a stunningly vast and profound study of the works of the early Churchfathers, who, in order to refute them, quoted extensively and relatively trustworthy from pagan philosophical sources, especially the Stoic ones that determined to a large extend the intellectual life of their day.[184] They had developed the concept he needed, but in un unacceptable form. Newton was furthermore familiar with the work of the great Renaissance scholars that compiled, edited and commented on these Stoic sources.[185] The General Scholium is a case in point for the combined, and

180 Iliffe, *Priest*, 104-105.
181 *Theological Notebook* (Keynes MS 2). An electronic edition is available on the Newton Project.
182 "Two notable corruptions", "Paradoxical questions regarding Athanasius" (N563M3 P222, William Andrews Clark Memorial Library, Los Angeles), "Irenicum" (Keynes 3). All are on The Newton Project.
183 REFS Newton's teological studies
184 The first part of Keynes Ms 2, the "Theological notebook" contains at the start an endless list of Patristic sources Newton consulted. See The Newton Project. Also Mandelbrote, "Fathers".
185 "Here is again Lipsius's connous intention to harmonize the ideas of eminent ancient sages. This kind of operation was frequent in the Renaissance n the basis of the belief in the *prisca theologia*", Hirai, *world soul*, 72.

profound, effects these two influences — philosophical and theological — had in shaping Newton's basic intellectual framework. I therefore agree with Levitin when he states,

> Understanding that for Newton the history of philosophy has to be read in the context of the history of religion is the key that lets us understand both the Classical Scholia and the relevant parts of the General Scholium and the Queries to the Opticks.[186]

Building on this double foundation, we investigated in earlier work the transmission of Stoic influences through Renaissance and Early Modern philosophers like Lipsius on Newton.[187] Studies detailing such Stoic influences remain relatively rare, however, and there has been a tendency to downplay its importance. After all, if the Stoics are so important, why does Newton say so little about them? Indeed, the Classical Scholia do not contain a single reference to a Stoic philosopher *qua* Stoic philosopher; when names with clearly Stoic affiliations are mentioned, like Cicero's, it is almost always to impart information about someone else.[188] Newton would have been prepared to quote even Lucretius verbatim in his scholia to propositions VI and VII. So do we see something we only *want* to see, or is there something else going on?

It should be kept in mind that "Stoicism" is not a homogenous entity. The conceptions of the early Stoics (Zeno, Cleanthes) differ in important aspects from the later doctrines.[189] Chrysippus introduces the already existing notion of a cosmic *pneuma* (Latin: *spiritus* or *aether*) to resolve certain inconsistencies which bothered the earlier theories about the cosmic creative fire (πῦρ τεχωικόν), because fire also existed as one of the four *stoicheia* or elements. They nevertheless all have a notion of an all-pervading material substance[190] in common that binds the cosmic fabric together. It is not just one of the four elements, as is clear from the fact that the Romans translated it systematically as *spiritus*, not *aer*.[191] Chrysippus's démarche is so succesful that it at once constitutes the summit and the creative end of Stoic cosmology — for centuries to come, people will broaden the scope of this concept into moral, social and theological directions without essentially changing it. This aether/pneuma is an "animal"[192] — a living soul, and is equated to *theos* — God. "The vis anima divina is the Stoic πνεῦμα: it 'breathes' through the universe." [193] and is its omnipresent "command centre" (hegemonikon, τὸ ἡγεμονικὸν). This state of affairs evidently opens up many possibilities for early Christian thinkers, who see Stoicism as a means to make sense of monotheism for a cultured Graeco-Roman public, and thus an ideal tool in their efforts to convert the remaining pagans.[194] Extensive treatments of the different philosophical schools in the early Churchfathers thus served two goals simultanously: philosophical reinforcement of existing Christian theological doctrine and proselytisation.[195] But the materialist side of this doctrine posed a serious theological problem. Evidently, an adaptation of the Stoic doctrines to the core requirements of Christian dogma was necessary to achieve these goals, and it was a matter of constant and vivid controversy and debate.[196]

---

[186] Levitin, *Ancient Wisdom*, p. 439.
[187] De Smet and Verelst, "Platonic and Stoic Legacy".
[188] Altough Cicero's importance as a transmitter and commentator of Stoic ideas could hardly be overstated: "Cicero holds a special place in the history of the Stoic tradition, a place second only to that occupied by the middle and Roman Stoics themselves." Colish, *Stoic Tradition I* (Leiden 1985): 61.
[189] M. Lapidge, "ἀρχαί and στοιχεῖα: A Problem in Stoic Cosmology", *Phronesis*, 18, 3 (1973): 240-278.
[190] Evidently not in the Peripathetic sense. "Aristotle's 'first matter' was inseparable, incorporeal and knowable only by analogy." Lapidge, "Stoic Cosmology", 244.
[191] Lapidge, "Stoic Cosmology", 274.

Stoic doctrine is deeply problematic from the point of view of both Newton's philosophical and his theological stance as well. Yet at the same time it provided him with a key concept, the cosmic *spiritus* that pervades and animates everything and binds the world together, a concept that he needs to bridge the causal gap in his natural philosophy and the ontological one in his theology. How does gravity, as force necessarily immaterial[197], causally interact with things in the material world? And how does God

---

[192] See Cicero's discussion of Stoic ideas in his *Academica*, e.g. in book xxxvii, 119, a passage Newton refers to (Schüller, *Newton's Scholia*, p. 230). For the older doctrines, see Hahm, *Stoic Cosmology*, 140-141. This pernicious notion is discussed by Lipsius with reference to, a.o., Plato's *Timaeus* and Posidonius. The latter one is among the references Newton rejected in the passage from Macrobius, discussed before (ft. 145). He does keep the reference to Plato's *Timaeus*. Intriguingly, in the ultimately rejected *Conclusio* he wrote for the *Principia* in 1687, Newton critically mentions the *spiritus animales (quos fingunt)* (**"the animal spirits which they feign"**) as incapable of adequately explaining muscular bodily movement, in the context of a discussion on the nature of attractive forces. "They" refers to the Cartesians ("Philosophia vulgaris"). Newton adds, "I am far from affirming that my views are correct, and I acknowledge their great imperfection, nevertheless they are simple and easy to conceive, and **of the same kind as the natural philosophy of the cosmic system**" (bold added). The parallel to the final pararaph of the General Scholium decades later will be obvious. This proves that the notion of "spiritus" plays a key rôle right from the start, and it a least to some extend explains Newton's caution when quoting ancient sources dwelling on it. See Hall and Hall, *Unpublished Scientific Papers*, 332-333; 345-346.

[193] M. Lapidge, "A Stoic Metaphor in Late Latin Poetry: the Binding of the Cosmos", *Latomus* 39, 4 (1980): 820. —, "Stoic Cosmology", 164 169

[194] M. Frede, "Monotheism and Pagan Philosophy in Late Antiquity", in: P. Athanassiadi / M. Frede (Eds..), *Pagan Monotheism in Late Antiquity* (Oxford, 1999): 41-67.

[195] M.L. Colish, *The Stoic Tradition from Antiquity to the Early Middle Ages II. Stoicism in Christian Latin Thought through the Sixth Century*, Brill (Leiden, 1985).

[196] Colish, Stoic Tradition II.

[197] As he concluded after many attempts to reconcile some form of mechanical philosophy with the experimental data — no 'drag' that causes the planets to diverge from their almost perfectly Keplerian paths around the Sun. Dobbs, *Janus Faces*, 185-191. In this respect, Cotes famous introduction to the second edition of the *Principia* is worth considering, because it gives a detailed account of Newton's arguments

interact directly and unmediatedly with His creation?[198] The crucial insight concerns, of course, how these questions are related — and this is precisely *the* topic dealt with in the General Scholium. Newton's theology was extremely sensitive to the precise interpretation of such notions because of their implications for his natural philosophy, but also because of his heterodox, anti-Trinitarian theological stance that would be irreconcilable with whatever "metaphysical (substantial) explanation" of God's nature.[199] For Newton, the Stoics are like an ancient version of the modern Cartesians and, like Descartes's philosophy, they are everywhere in his thought, even though they are virtually never mentioned. They deal with all the issues he is dealing with, but they provide a solution which he cannot accept. He hates them and yet he needs them, if only as a way to define what his philosophy is not. The General Scholium tells us that in no uncertain terms, while it simultaneously shows us that Newton looked at the Stoics for inspiration, since their *spiritus* appears expressly at the end of it.[200]

These almost contradictory requirements explain why Newton handled the Neostoic scholars of his day with such utter caution. It also explains their quasi invisibility in his own work. In this contribution I shall theerefore take another approach to unearh the Stoic influence, by looking closer at the ways in which Stoic authors of Antiquity — whether Christian[201] or pagan — shaped Newton's ideas directly. Newton got much of against a mechanical aether. Edition Cotes Newton' Newton's ideas with respect to the aether evovled over time and remained always somewhat ambiguous. Gravity is an interaction (it decreases with distance), not a quality, it can be measured and obeys mathematical laws, but is not some kind of material entity or property. It pervades bodies without mechaniclly affecting them. It acts as a medium, but in a non-mechanical manner. For a clear explanation of these subtleties, see Janiak, *Philosopher*, 93-101.

198 Without having recourse to innate occult qualities in the sense of Leibniz. This is precisely why not merely divine presence, but also divine action is required, as Newton explains in the famous letter to Bentley (25 february 1693). Turnbull, *Correspondence* 3, 253-254. Compare to Gregory in his Memorandum of 5,6,7 May 1694. "[Newton says] that a continual miracle is needed to prevent the Sun and the fixed stars from rushing together through gravity: that the great eccentricity in comets in directions both different from and contrary to the planets indicates a divine hand: and implies that the comets are destined for a use other than that of the planets. The Satellites of jupiter ans Saturn can take the places of the Earth, Venus, Mars if they are destroyed, and be held in reserve for a new Creation." *Correspondence* 3, 336-337

199 R.S. Westfall, *Never at Rest*, Cambridge UP, 1980, 309-334; J.E. McGuire, "The Fate of the Date: The Theology of Newton's Principia Revisited", in M. Osler (ed.), *Rethinking the Scientific Revolution* (Cambridge, 2000): 271-297; S. Snobelen, "Isaac Newton, heretic: the strategies of a Nicodemite", *British Journal for the History of Science,* 32 (1999): 381-419.

200 It would lead us too far to go into the many links that exist with the Queries to the *Opticks*, especially when we take the draft material for it into account. McGuire and Rattansi, "Pipes of Pan", 118; Dobbs, *Janus Faces*, p. 197; Turnbull, *Correspondence* 3, 339 ft.10; Cohen, *Introduction*, 195, Ducheyne, "General Scholium".

201 Calling Christians "Stoics" may raise some eyebrows, but Colish criticises methinks rightfully the "scholarly myopia" that refuses to take the strongly Stoicizing tendencies in some of these early Christian writers seriously, and looks at them only from the point of view of "classical bias"; as mere preservers and transmitters of the older ideas, not as thinkers in their own right. Colish, *Stoic Tradition II*, 1-2. The inherentaly inconsistent attitude many of them entertain to Stoicism reflects the delicacy of the required intellectual exercise. "Our own investigation of Stoicism in Tertullian makes it impossible to sustain the his information from patristic sources, Churchfathers like Eusebius, Origines, Lactantius, Clemens.[202] It is not my intention here to discuss in any depth the precise philosophical and theological reasons for, the implications of and argumentative strategies in which Newton used such-and-such an ancient author. To do so properly would require a monograph. I shall limit myself to uncovering some direct interconnections by form and content between source-texts and Newton's fragments in as far as they deal with the *spiritus*, and comment briefly on the larger intellectual framework in which they fit. The reader does wise to keep Dobbs's summary in mind, according to which Newton conflated "Thales with St. Paul by way of the Stoics"[203] — and to realise that what we find back in the textfragments — especially those potentially destined for public communication — is often only Thales and St. Paul, the Stoic middle term having suddenly disappeared, while being essential to the correct conclusion and understanding of the overall argument.

## "From Thales to Paul through the Stoics."

The Stoic notion of interest to Newton, "spiritus", had been introduced into Christianity by an early Latin Christian writer, Tertullian, who brought materialistic Stoic philosophy in the field as a weapon against Gnostic dualistic idealism. His main focus in tis context is the nature of God and His relation to the universe.[204] He distinguished the *pneuma* of God ("spiritus") from the *pneuma* of the human soul ("flatus") as being of a different kind of materiality, but a similar kind of activity, namely, as the principle of unity, the seat of life and of free will.[205] Through this materiality, God can create: the Father brings forth the Son and the Son brings forth the Holy Spirit; they are different but share a portion of this divine materiality — Tertullian is also the first to introduce the notion of Trinity into Christianity, as the *tres personas, una substantia*-doctrine: not yet in the later sense of the "Triune God"[206], but nevertheless the starting point for what will become the Nicene Creed. In one stroke we see all the theological topics that will occupy Newton throughout his lifetime appear. The intimate link between Stoic and Trinitarian concepts, like

---

view that he is primarily or exclusively a supporter, an enemy, or a transformer of Stoicism. He does all of these things simultaneously and to approximately the same degree." *id.*, 13.

[202] Cfr. ft 180.

[203] Dobbs, *Janus Faces*, 115.

[204] Colish, *Stoic Tradition II*, 15.

[205] G. Verbeke, *L'évolution de la doctrine du pneuma*, Desclée De Brouwer, (Paris, 1945): 442-448.

[206] D. Tuggy, "Trinity", *The Stanford Encyclopedia of Philosophy* (Winter 2016 Edition), Edward N. Zalta (ed.), URL = <https://plato.stanford.edu/archives/win2016/entries/trinity/>

*homoousios*[207], the consubstantiality between the Father and the Son, henceforth occupy the stage of the intellectual debate.

Interest in ancient philosophical schools as a weapon in theological battles is rife at the beginning of the Modern era[208], and accounts largely for the revival of interest in ancient Stoicism during the sixteenth century in the so-called Neostoic school. In the context of the heated contra-reformation debates on the dogma of transsubstantiation, the Trinity[209] and the nature of the resurrected body, monistic theories on the potentially material effects of God's presence in His creation were looked upon with utmost suspicion, so there is again not much of a surprise in finding Newton to be extremely careful with his handling of such material. We showed in earlier work nevertheless that Neostoic influences were actively working on Newton while writing his General Scholium. The fact that almost all classical references Newton uses in his gloss to the General Scholium are already exactly so present in Justus Lipsius is just one case in point.[210] Newton, as is by now well known, was an active all be it extremely discrete participant in those debates.[211] That is one of the reasons why references to Stoic sources not only in published, but even in private versions of his work are rarely explicit, often hidden behind indirect references or obscured in quotes shared by several authors or containing a plurality of references themselves. Interestingly, in the Classical Scholia there is at first glance very little reference to Stoic philosophy even though it is plain and clear that Newton knew the material very well. The only obvious exceptions are the (not incidentially) poets Aratus and Vergilius. Aratus can explicity be referred to the Stoic intellectual environment of his time (he was an eclectic who at some point was a student of Zeno of Citium).[212] And the Stoic influences on Virgil are well attested.[213] Indeed, more than just a standard-bearer for Lucretian atomism, Virgil has been also firmly associated with Stoicism since Antiquity. Newton's citational strategy fits itself in "a citational practice that is specific to the philosophical school of Stoicism: the mining of extant literature for evidence of specific doctrines"[214] — in his case, for versions of doctrines of the *spiritus* not only fitting his philosophical agenda, but also, and crucially, his theological aims. It allows us again to see Newton's two-layered method of reading ancient texts at work, tapping into Stoic ideas through indirect and direct sources, but concealing their visible presence[215] while recuperating a reformed version of some of their most basic ideas.

---

207 Iliffe, *Priest*, 144-149.
208 For context relevant to Newton, see Levitin, *Ancient Wisdom*; Iliffe, *Priest.*
209 Iliffe, paper on Erasmus
210 De Smet and Verelst, "Platonic and Stoic Legacy", 16.
211 Iliffe, *Priest*, 123. Snobelen, "Discourse", on Newton's use of "the Principle of Obliquity".
212 H.C. Baird, "Stoicism in the Stars. Cicero's Aratea in De Natura Deorum", *Latomus*, 77, 3 (2018): 646-670. See also Dobbs, *Janus Faces*, 200.
213 Colish, *Stoic Tradition I*, 225: "Vergil displays a profound and perceptive feeling for the ethical and metaphysical implications of Stoic philosophy. To the extent that he appeals to Stoic principles in the *Aeneid* he internalizes them thoroughly and weaves them seamlessly into the web of his epic. Yet, he subordinates them ultimately to his own personal poetic vision." Also M.W. Edwards, "The expression of Stoic Ideas in the 'Aeneid'", *Phoenix*, 14, 3 (1960): 151-165.

Newton paves the way for us himself. At the end of the scholium to prop. IX, he says that the "older philosophers" saw the world as a temple, the body of God with a fire in the middle that binds everything together. He also refers to Pan, whose play with the harmonic ratio binds the souls of the world through numbers and connects this to his gravitational force. He then cites Macrobius's comment on Cicero, who quotes Virgil:

> The same view of the philosophers was related by Virgil: for he, too awarded the world a soul and, in order to attest its purity, he called it mind.

Then the passages out of Macrobius's comments on Cicero's *Dream of Scipio* which we discussed before take a start. This piece is Cicero's philosophical tour de force, and it is thoroughly Stoic in nature.[216] Newton will have to do some philosophical clean up once again to make it fit his stall, and that is as we saw before precisely what happens: he follows closely the truncated citations from Virgil's poem discussed in Lipsius[217], and then goes on to selectively quote from Macrobius where he leaves out precisely those philosophers that support a too material interpretation of the world soul.[218] Newton basically builds his alternative ancient analogue for the Stoic/Cartesian material aether by digging up fragments on an immaterial, mathematical principle that can take on the binding, harmonising rôle of the *spiritus*: caelestial music, the harmony of he spheres. The Stoic guides the non-Stoic, so to say.

There are other, even more telling examples. I mentioned before that the 1713-version of the General Scholium[219] contains a gloss to *ipso* in the sentence:

---

214Baird, "Stoicism in the Stars". Baird discusses in detail the use Cicero makes of this technique in his translation of Aratus's *Phaenomena*, in which "the criteria for inclusion and omission of specific verses were guided by Stoic principles." Newton refers to both works, and uses a similar techique to either enhance, either conceal aspects of their docrines as it befits him.
215 Using what Snobelen calls "the principle of obliquity", ""To discourse of God',
216 "The dream of Scipio at the end of the *De republica* develops to its most explicit level Cicero's correlation between statesmanship and Stoic ethics and cosmology." Colish, *Stoic Tradition I*, 94.
217 Lipius, *Physiologia stoicorum* VII, 15. Cfr. Hirai, *world soul*, 67-68.
218 Cfr. ft. 162.
219 *variorum*-edition, 762.

> In ipso continentur & moventur universa, sed sine mutua passione. (In him are all things contained and moved; yet neither affects the other)[220]

This sentence conveys in a terse, but multilayered way the relation between God and absolute space, the container of everything — except God. Let us start by taking a look at

Newton's references to classic authorities in this gloss:

Cicero *De natura deorum*, lib 1 (on Pythagoras, Thales and Anaxagoras)
Virgil, *Georgica*, lib iv.v & Aneid. lib. 6. v. 721
Philo *Legum allegoriae* lib. 1. sub initio
Aratus *In Phaenomenis*, sub initio

Paul, Acts 17:28

Take a good look again. All the cited non-Christian authors — literally ALL of them — are renowned Stoics or at least have strong ties to Stoicism, but NONE of them is staged in that capacity. Cicero was a critical sympathiser of the (moral philosophy of) the Stoa, but is quoted here in his capacity as scholar, not as a philosopher in his own right. Virgil's poetic descriptions of the *anima mundi* in all of his major works have been taken since Antiquity for what they are: largely Stoic in inspiration. But Virgil is a poet, not a philosopher. Philo is a notorious philosopher, the key figure indeed in the transfer of Stoic ideas into the Jewish intellectual community of the centrum of learning which was Alexandria in his day, but he was a Jew, and so belonged formally to a different tradition. And finally the Hellenistic poet Aratus, who had been a student at the school of Zeno of Citium is, well, a poet, not a philosopher.

The passages from Cicero give a cursory overview (and a quite biting critique) of the ideas of the Ionic philosophers, to the extend that they all had some kind of element or principle that they put at the foundation of the world. The impression is that Newton, who definitedly had more reliable sources at his disposal, did not care too much about accuracy at this point, and just wanted to make the point of their rôle as precursors.[221]

---

[220] *The Principia*, Transl. Motte-Cajori, 441.
[221] In the Classical Scholia, much more work goes into doing justice to "Ionic philosophers" like Pythagoras and Thales, e.g., in the scholia to prop. VI, IX.

This squares with the fact that not all drafts of the General Scholium (in fact, a minority) do carry this reference.[222]

On Virgil's all-pervading world soul (*aether*) I said enough above. Let me just add the verses out of Georgica, that make the same point but transposed to the world of the bees:

**Georgica, Liber IV, 219-227**

His quidam signis atque haec exempla secuti
esse apibus partem divinae mentis et haustus
aetherios dixere; deum namque ire per omnes
terrasque tractusque maris caelumque profundum.

Noting these tokens and examples some have said
that a share of divine intelligence is in bees,
and a draught of aether*:* since there is a god in everything,
earth and the expanse of sea and the sky's depths:

Philo is present not only in the gloss, but also in the wording of this part of the General Scholium.[223] To summarise, in the intended passage, Philo says that "Deus implet omnia, penetrat omnia. (...) Et quem locum occupabis, in quo non sit Deus?" Philo "stoicises" doctrines that were familiar to rabbinic theology, like the doctrine of God's omnipresence.[224] In biblical cosmology, God dwells in the heavens and makes them holy by his presence. Nothing in the world is so lowly that God would not bless it by dwelling in it. Solomon, praising God, said: "Heaven itself, the highest heaven, cannot contain thee" **1 Kings 8 : 27**. And God says in **Jeremiah 23 : 24** that "He fills heaven and Earth". At some point the Hebrew word for 'place' (*mãqôm*)[225] simply becomes a name for God.[226] R. Ammi in a comment on Genesis asks:

"Why do we give a changed name to the Holy One, blessed be He, and call him 'the Place'?"

and gives himself the answer:

---

[222] Cfr. Snobelen's list of General Scholium versions

[223] De Smet and Verelst, "Platonic and Stoic Legacy", 8. For an interesting comment on the relation between 'mind' and 'heaven' in this context, see Dobbs, *Janus Faces*: 200-201.

[224] B.P. Copenhaver, "Jewish Theologies of space in the Scientific Revolution: Henry More, Joseph Raphson, Isaac Newton and their predecessors", *Annals of Science* 37, 5: 489-548; 491.

[225] Cfr. McGuire, "Newton on Place, Time and God", pp. 120-121.

[226] A. Ramati, "The hidden truth of creation: Newton's method of fluxions", *British Journal for the History of Science* 34, 4 (2002): 417-438, especially 427-428 and ft. 54.

"Because He is the Place of the world."[227]

Finally Aratus. Newton quotes his *Phaenomena* in the Classical Scholia, as part of a series of excerpta from — again — Macrobius's *Dream of Scipio*:

> ipsum denique Jovem veteres vocaverunt et apud Theologus Juppiter est mundi anima. hinc illud est
>
> Ab Jove principium Musae, Jovis omnia plena,
>
> quod de Arato poetae alij mutuati sunt
>
> The ancients called it Jupiter, and to theologians Jupiter is the soul of the world. Therefore is written
>
> With Jove I begin, ye Muses; Of Jove all things are full
>
> which the other poets borrowed from Aratus[228]

Intriguingly, the last part of the verse quoted by Macrobius is not Aratus's, but Virgil's![229] This is probably one of the "other poets" Newton had in mind when he makes his final remark. Virgil's verse stems from the Bucolica/Eclogues III, 60:[230]

> Ab Iove principium, Musae: lovis omnia plena;
> iIle colit terras, iIli mea carmina curae.
>
> With Jove I begin, ye Muses; of Jove all things
> are full. He makes the earth fruitful; he pays heed
> to my songs.

---

[227] Copenhaver, "Jewish Theologies of space", 493. There is a link to the Patristic tradition as well. After citing Jeremiah 23 : 24 and several other texts in support of omnipresence, Gregory concludes with a dithyramb on God's power to "inhabit, penetrate, surround, support, and rule the universe". Arnobius, *Adversus Nationes* 1.31, contains a prayer in praise of God which calls him "... the First Cause, the place and true space, the foundation of all things that exist, the only infinite, Unbegotten, Immortal, Eternal, whose lineaments no bodily shape contains...", which alludes to a line of Lucretius and represent the location "... in which the world that contains us ... sits and turns". Arnobius, in *Adversus Nationes* 1.31 has a link to Lucretius: "Prima enim tu causa es, locus verum ac spatium, fundamentum cunctorum quaecumque sunt..." Lucretius *Rer. nat.* 1.471-472: "denique materies si rerum nulla fuisset, nec locus ac spatium, res in quo quaequa geruntur...". Copenhaver, "Jewish Theologies of space", 497-498. Cfr. McGuire, "Newton on Place, Time and God", pp. 120-121.

[228] Transl. Schüller, *Newton's Scholia*, p. 245.

[229] Clemens cites Aratus's last verse in a contextus in which the Stoics (and Pythagoras) are explicitly mentioned. I quote from the latin translation in the Patrologia Latina: "Qui solus, inquit, illucescente die, noctem efficere potest, Deus est. Et Aratus in Phaenomenis haec habet: Ab Jove principium, nullo qui tempore nobis praetereundus erit. Cujus late omnia numen Compita, et omne forum, pontique profunda tumentis, Atque omnes portus implet: nemoque potenti Auxilio caruisse queat." In *Opera quae extant*, PG 8, "Cohortatio ad Gentes", Cap. V, col 151 B,C.

[230] Fairclough, *Virgil*, Loeb Classical Library, vol. I, pp. 20-21.

But is it interesting for yet another reason: Virgil was not the only one to "borrow" Aratus's verse. This brings us to the only reference in the General Scholium-gloss to a Christian author, **Paul, Acts 17:28**. The crucial sentence in that bible verse is:

> For in him we live, and move, and have our being,

This is literally the sentence we find back in the General Scholium. And again in Lipsius's *Physiologiae stoicorum* as well.[231] The latin text in the Vulgate-edition of the Bible[232]:

> **In ipso enim vivimus, et movemur**, et sumus: sicut et quidam vestrorum poëtarum dixerunt: Ipsius enim et genus sumus (**For in him we live, and move, and have our being**; as certain also of your own poets have said, for we are also his offspring).

This quote is part of a larger speech that Paul famously held on the Areopagus in Athens in A.D. 51[233], a city which he respects for her culture and learning, but he is appaled by the widespread "idolatry" — people worshipping "an unknown god" by means of statues and temples. Paul starts to preach the gospel and is somehow invited to come to the Areopagus — not far from where the Stoics hold school — to explain himself to Stoic and Epicurean philosophers and everybody else who might be interested, at least if we can believe the report Luke gives us.[234] Paul is a learned man and gives a brillaint speech in which he he addresses both the commoner and the philosopher[235], which explains the very Stoic vocabulary and tone of parts of his address.[236] Not only the way he describes the — uncharacteristic for the Greek mind — personal God in terms reminiscent of the all-pervading Stoic spirit, but he also cites one of the most famous Hellenistic poets, Aratus, "as some of your own poets have said,***for we are also his offspring***").

> Aratus, *Phaenomena* 1:

---

[231] Dobbs, *Janus Faces*, p. 204-205, especially ft. 109. De Smet and Verelst, "Platonic and Stoic Legacy", 16.
[232]https://www.logosapostolic.org/bibles/latin_vulgate_textus_receptus_king_james/acts/acts17.htm
[233] C.J. Hemer, "The Speeches of Act. II. The Areopagus Address", *Tyndale bulletin* 40,2, 1989, pp. 239-259.
[234] Archeological and literary data confirm the vibrancy of different philosophical schools in the Athens from the Hellenistic to the Roman period. M. Haake, "Philosophical Schools in Athenian Society from the Fourth to the First Century BC: An Overview", *Private Associations and the Public Sphere, Proceedings of a Symposium held at the Royal Danish Academy of Sciences and Letters*, 9-11 September 2010: 57-91. J. McK Camp, "The Philosophical Schools of Roman Athens", *Bulletin Supplement (University of London. Institute of Classical Studies)* 55, *The Greek Renaissance in the Roman Empire: Papers from the Tenth British Museum Classical Colloquium* (Oxford, 1989): 50-55.
[235] P. Gray, "Implied Audiences in the Areopagus Narrative", *Tyndale bulletin* 55,2 (2004): 205-218.
[236] M. Dibelius, "Paul on the Areopagus", in *Studies in the Acts of the Apostles*, London (1956): 63.

> From Zeus let us begin; him do we mortals never leave unnamed; full of Zeus are all the streets and all the market-places of men; full is the sea and the havens thereof; always we all have need of Zeus. ***For we are also his offspring***; and he in his kindness unto men giveth favourable signs and wakeneth the people to work, reminding them of livelihood.

The speech became somewhat of a *cause célèbre*, and the sentence "According to the Poet cited by the Apostle"[237] a catchphrase in the religiously inspired literary tradition. Dobbs, for once, saw the fish pass by, but did not catch it:

> The famous Pauline statement on God as the ground of all being in acts 17: 27-28 was probably drawn from the prologue of Aratus. (...) Newton seemed to indicate his knowledge of that filiation in his draft comment of the 1690 when he mentioned "the Poet cited by the Apostle". In the *Principia*'s General Scholium, of course, Newton cited the Prologue of Aratus directly.[238]

The "1690s draft" Dobbs is referring to is probably the manuscript partially published by McGuire and Rattansi, in which "the immaterial, 'immechanical' cause of [gravity], is seen to be God himself (…) There is little doubt that Newton saw in analogy to musical harmony, the principles of law and order in the natural world. Such harmony was the profoundest expression of cosmos (…) the providential governance of a Divine power actually present in the world". [239]

> (...) those ancients who more rightly held unimpaired the mystical philosophy as **Thales and the Stoics**[240], thought that a certain infinite spirit pervades all space into infinity, and contains and vivifies the entire world. And this spirit was their supreme divinity, **according to the Poet cited by the Apostle. In him we live and move and have our being.**

Here Newton drops all masks and throws his cards on the table: the (transformed) Stoic *anima mundi* is the secret connection between the ancient and the Biblical world both of the Old and New Testament. Indeed, in this light the Newtonian references to Old Testamentic sources like King's and Jeremiah acquire as well a new meaning. Thus in the

---

[237] R. Faber, "The Apostle and the Poet: Paul and Aratus", *Clarion* 42, 13 (1993). https://spindleworks.com/library/rfaber/aratus.htm
[238] Dobbs, *Janus Faces*, 200.
[239] CUL Add. 3965.12, f269. Bold added. McGuire and Rattansi, "Pipes of Pan", 120, ft. 25. For other fragments referring to the Apostle, see McGuire, "Newton on Place", 120-121.
[240] i.e., basically the ancient authorities referenced in the gloss to the published version of the GS.

General Scholium, the Poet and the Apostle meet the Natural Philosopher on the crossroad of two texts and two traditions. Aratus and Paul reach out to each other, and Newton is the one who through his natural philosophy finally fulfills the unfulfilled promise at the end of Antiquity by opening the minds and hearts of modern man again to true knowledge and understanding, of nature, and through nature, of God.

Newton evidently transforms profoundly the Stoic ideas he borrows and it does seem likely that keeping the rôle those Stoics played in his own framework out of sight was his deliberate choice. Consequently his somewhat bizarre method of referencing his sources does reflect a consciously designed strategy and appears to be rational from his point of view; it moreover also fits within an already existing tradtion.[241] Dobbs speaks about Newton's "post-*Principia* spiritual *pneuma*".[242] The very strong impression exists that, rather than obliging himself to engage in a metaphysical discussion with the philosophical proponents of materialistic schools like the Cartesians or the (neo)Stoa, — we know how "fond" he was of that kind of exchange — Newton chooses to to use their lucky expressions of a "spiritus" or "anima mundi" merely as a 'figure' for transmitting his idea of gravity, a metaphor of celestial music produced by a divine and continuous spirit ("continuatus spiritus"[243]) that binds the cosmos together immaterially in an ordered whole.[244] The fact that Newton held "pondere, numero et mensura" as a personal motto then comes in as somewhat less than a surprise.[245] Given what he actually does, it sounds somewhat ironical to hear Newton state at the end of the General Scholium that he would really like to say more, but that he cannot because he lacks the "copia experimentorum" to do so. Rhetorically speaking, the set up of the overall argument of the General Scholium is again a classic masterpiece...

---

241 With respect to Stoic citation: Baird, "Stoicism in the Stars". For Newton's strategy, Snobelen, *Discourse*.
242 Dobbs, *Janus Faces,* 199; Also Janiak, *Philoosopher*, 100, ft. 21.
243 Cicero, *De natura deorum* II.19. See Lapidge, "A Stoic Metaphor", 817-837.
244 Dobbs, *Janus Faces*, 196.
245 Dobs, *Janus Faces,* 155. Snobelen, PNEM-paper. Cfr. ft 56.

APPENDIX

RS MS 247 — CS missing page 14r
[Edidit Craufurd, 1829][246]

Sigla:

[???] unreadable deletion
s.l.: supra linea
——: deletion, omission
Abbreviations are solved between []

~~Perfectiss~~[247] Deum esse Ens summe[248] perfect[~~issi~~]um concedunt omnes. Entis autem summe perfecti[~~issimi~~] Idea est ut sit substantia una, simplex, indivisibilis, viva et vivifica, ubiq[ue] semper necessario existens, summe intelligens omnia, libere volens bona, voluntate efficiens possibilia, effectibus nobilioribus similitudinem propriam quantum fieri potest communicans, omnia in se continens tanquam eorum principium et locus, omnia per praesentiam ~~suam~~ substantialem[249] cernens et regens (sicut hominis pars cogitans sentit species rerum ~~qua~~ in cerebrum ~~diferuntur~~ delatas et illinc regit corpus proprium,) & cum rebus ~~alijs~~ omnibus[250] secundum leges accuratas ut naturae totius fundamentum & causa[251] constanter cooperans, nisi ubi aliter agere bonum est. ~~et rationi consentaneum~~ Nam liberrime agit quae optima & rationi maximè consentanea sunt, et errore vel lato caeco adduci non potest ut aliter agat. Haec est Idea Entis summe perfecti. Et conceptus durior Deitatem minime perficiat sed suspectam potius reddet, aut forsan[252] excludet e rerum natura.

Quicquid necessario existit illud semper et ubiq[ue] existit, cùm eadem sit necessitatis lex in locis et temporibus universis. Et hinc omnis rerum diversitas quae in locis et temporibus diversis reperitur ex necessitate caeca non fuit sed a voluntate entis necessario existentis originem duxit. Solum enim ens intelligens vi voluntatis suae secundum intellectuales rerum ideas[253] propter causas finales agendo varietatem rerum introducere potuit. Varietas autem

246 Craufurd, "Notice".
247 s.l.
248 s.l.
249 s.l.
250 s.l.
251 ut... causa s.l.
252 aut forsan s.l.
253 secundum...ideas s.l.

in corporibus maxime reperitur & corpora quae in sensus incurrunt sunt Stellae Fixae, Planetae, et Cometae, [???] Terra et eorum partes.

Coelos et spatium universum aliquà materia fluida sublissima implent, sed cujus existentia nec sensibus patet nec ullis argumentis convincitur, sed ~~praecariò~~ hypotheseos alicujus gratia praecario assumitur. Quinimo si et rationi ~~et sensibus,~~ fidendum sit et sensibus materia illa e rerum natura exulabit. Nam quomodo motus in pleno peragatur [~~nemo ???~~] intelligi non potest; cùm partes materiae, utcunq[ue] minutae, si globulares sint, nunquam implebunt spatium solidum; sin angulares ~~inter se~~ ~~firmius haerebunt quam la~~ propter omnimodum superficierum contactum firmius haerebunt inter se firm quam lapides in acervo, et ordine semel turbato non amplius ~~spatium solid~~ congruent ad spatium solidum implendum. Porrò tam experimentis probavimus quàm rationibus mathematicis[254] quod [???] resistentia quam cor~~pora in fluidis progredientia sentiunt sit ut [motus]~~ ~~fluidi densitas & quod in fluidis quadratum velocitatis conjunctim & quod in fluido ejus dem densitatis cum corpore moto, corpus sphaericum[255] pro~~ ~~describendo longitudinem diametri suae amittat plusquam semissem motus sui pus in fluido [cor]pus sphaericum[256] cujuscunq[u]e~~ densitatis cujuscun[que][257] in fluido ejusdem densitatis utcunc[que] subtili[258] progrediens, ~~describendo longitudinem diametri suae~~ [???] ex resistentia medij prius[259] admittet semissem motus sui quam longitudinem diametri suae descripserit. Et quod[260] resistentia fluidi illius nec[261] per subtilem partium divisionem [~~vel~~] nec per motum ~~parvum~~ partium[262] inter se diminui [~~potest~~] possit ut corpus ~~minus resistatur quam pro~~ longitudinem diametri prius describat quam amittat semissem motus.

---

[254] s.l.
[255] s.l.
[256] s.l.
[257] s.l.
[258] utcuncque subtili s.l.
[259] s.l.
[260] s.l.
[261] illius nec s.l.
[262] s.l.